\documentclass[showpacs,aps,prd,nofootinbib,floatfix,amsmath,amssymb]{revtex4-2}
\usepackage{graphicx}
\usepackage[utf8x]{inputenc}
\usepackage{color}
\usepackage{subfig}
\usepackage{multirow}
\usepackage{slashed}
\usepackage{cancel}

\begin{document}

\title{Revisiting the top quark chromomagnetic dipole moment in the SM}
\author{J. I. Aranda$^{1}$}
\author{T. Cisneros-P\'erez$^{1}$}
\author{J. Monta\~no$^{1,2}$}\email{jmontano@conacyt.mx}
\author{B. Quezadas-Vivian$^{1}$}
\author{F. Ram\'irez-Zavaleta$^{1}$}\email{feramirez@umich.mx}
\author{E. S. Tututi$^{1}$}
\affiliation{
$^{1}$Facultad de Ciencias F\'isico Matem\'aticas, Universidad Michoacana de San Nicol\'as de Hidalgo,
Av. Francisco J. M\'ugica s/n, 58060, Morelia, Michoac\'an, M\'exico.
\\
$^{2}$C\'atedras Conacyt, Av. Insurgentes Sur 1582, Col. Cr\'edito Constructor, Alc. Benito Ju\'arez, 03940, Ciudad de M\'exico, M\'exico.
}

\date{\today}

\begin{abstract}
We revisit the anomalous chromomagnetic dipole moment in the Standard Model and show that its triple gluon vertex contribution, with the on-shell gluon ($q^2=0$), generates an infrared divergent pole. Consequently, the chromomagnetic dipole should not be perturbatively evaluated at $q^2=0$. Focusing on this top quark anomaly, denoted as $\hat{\mu}_t(q^2)$, we compute it with the off-shell gluon with a large momentum transfer, just as the $\alpha_s(m_Z^2)$ convention scale, for both spacelike $q^2=-m_Z^2$ and timelike $q^2=m_Z^2$ cases. We found that
$\hat{\mu}_t(-m_Z^2)$ $=$ $-0.0224$$-$$0.000925i$ and
$\hat{\mu}_t(m_Z^2)$ $=$ $-0.0133$$-$$0.0267i$. Our $\mathrm{Re}\thinspace\hat{\mu}_t(-m_Z^2)$ matches well with the current experimental value $\hat{\mu}_t^\mathrm{Exp}=-0.024_{-0.009}^{+0.013}(\mathrm{stat})_{-0.011}^{+0.016}(\mathrm{syst})$, and the $\mathrm{Im}\thinspace\hat{\mu}_t(-m_Z^2)$ part is induced by flavour changing charged currents.
\end{abstract}


\maketitle

\section{Introduction}
\label{sec:intro}

The top quark anomalous chromomagnetic dipole moment (CMDM) has been recently measured by the CMS Collaboration at the Large Hadron Collider (LHC) by using $pp$ collisions at the centre-of-mass energy of 13 TeV with an integrated luminosity of 35.9 fb$^{-1}$ \cite{Sirunyan:2019eyu}. Specifically, they reported
\begin{equation}\label{CMDM-experiment}
\hat{\mu}_t^\mathrm{Exp}=-0.024_{-0.009}^{+0.013}(\mathrm{stat})_{-0.011}^{+0.016}(\mathrm{syst}),
\end{equation}
whereas for the chromoelectric dipole moment (CEDM),
\begin{equation}\label{}
|\hat{d}_t^\mathrm{~Exp}|<0.03,
\end{equation}
at $95\%$ C. L.

In contrast, in the Standard Model (SM), the CMDM is induced at the one-loop level, and it receives contributions from both quantum chromodynamics (QCD) and electroweak (EW) sectors. A peculiar feature of this property concerns the existence of an infrared (IR) divergence generated by the Feynman diagram coming from the QCD non-Abelian triple gluon vertex. This issue occurs when the gluon momentum transfer $q$ of the external gluon is on-shell, $q^2=0$, which has been pointed out in Refs.~\cite{Choudhury:2014lna,Bermudez:2017bpx,Aranda:2018zis}. The authors in Ref. \cite{Choudhury:2014lna} were the first to show the presence of that IR divergence when the gluon is on-shell; they employed the Feynman parameterization (FP) method and realized that the corresponding calculation reported as finite in Ref. \cite{Martinez:2007qf}, through the same method, is incorrect. Nonetheless, this erroneous result has been considered the correct SM prediction by the community
\cite{Etesami:2018mqk,Aguilar-Saavedra:2018ggp,Hernandez-Juarez:2018uow,Etesami:2017ufk,Aguilar-Saavedra:2014iga}. Additionally, the same divergence issue was indicated in Ref. \cite{Bermudez:2017bpx} based on the integration-by-parts technique \cite{Davydychev:2000rt}.

In this work, we take the divergence discussion one step further: here, we will show, by using dimensional regularization (DR), the IR nature of that divergence by displaying its $1/\epsilon_\mathrm{IR}$ infrared pole
\cite{Kinoshita:1962ur,Bollini:1972ui,tHooft:1972tcz,Kinoshita:1975ie,Leibbrandt:1975dj,Dittmaier:2003bc,Collins:1984xc,Muta:2010xua,Ilisie:2016jta}, which comes from a two-point Passarino--Veltman scalar function (PaVe), identified as $B_0(q^2,0,0)$, when $q^2=0$. Consequently, in the context of perturbative QCD (pQCD), it is not suitable to evaluate the CMDM with the on-shell gluon: hence, in pQCD, it is not possible to establish a faithful analogy of the CMDM with the quantum electrodynamics (QED) static anomalous magnetic dipole moment defined with the on-shell photon $q^2=0$.
As a result, in Ref.~\cite{Choudhury:2014lna}, it was proposed to evaluate the CMDM at a large gluon momentum transfer, $q^2=-m_Z^2$. This choice is justified since in pQCD the strong running coupling constant is characterized at that conventional scale, $\alpha_s(Q^2=-q^2=m_Z^2)=0.1179$ \cite{PDG2020}, which depends on the momentum transfer $Q^2=-q^2$, where $q$ is the four-momentum flow of the process; it is established in the spacelike domain $Q^2>0$, implying $q^2<0$ \cite{Field:1989uq,Deur:2016tte,BeiglboCk:2006lfa,Nesterenko:2016pmx}.
Although the perturbative $\alpha_s$ is conceived in the spacelike regime $q^2<0$, its value is indistinctly used in the strong interaction processes, for example, in $t\bar{t}$ production at the LHC where the top quark chromodipoles are assumed in general as complex quantities in the timelike domain $q^2>0$ \cite{Bernreuther:2013aga,Khachatryan:2016xws}. Motivated by this fact, we will evaluate the $\hat{\mu}_t(q^2)$ at $q^2=-m_Z^2$ and at $q^2=m_Z^2$; in advance, we have found that both evaluations give rise to complex quantities, of which the $\mathrm{Re}\thinspace\hat{\mu}_t(\pm m_Z^2)$ parts are within the measured statistical error \cite{Sirunyan:2019eyu}, but in particular, our $\mathrm{Re}\thinspace\hat{\mu}_t(-m_Z^2)$ matches quite well with the central value. In Table III from Ref.~\cite{Aranda:2018zis}, some of us have already published our numerical results for the CMDM of the top quark in the SM at the $m_Z$ scale for both spacelike and timelike cases, $\hat{\mu}_t(q^2=\pm m_Z^2)$. However, in Ref.~\cite{Aranda:2018zis} the analytical details for the calculation of the CMDM in the SM were not presented, we discuss them in this work.

Another way to approach the IR divergence issue, which we also address in this work, is to apply the FP method \cite{Peskin:1995ev}, but without omitting the $+i\varepsilon$ Feynman prescription of the propagators; this is performed in order to analytically show that the gluon propagator in the Feynman-'t Hooft gauge $\xi=1$ and in the general renormalizable $R_\xi$ gauge leads to the same logarithmic IR divergence when $\varepsilon\to 0$. Furthermore, as carried out in Ref. \cite{Choudhury:2014lna}, we implement the massive gluon propagator artifice, but we do provide an analytical expression to prove that the logarithmic IR divergence arises when $m_g\to 0$.

The outline of this paper is as follows. Section \ref{Sec:CMDM-lagrangian} presents the general effective chromoelectromagnetic dipole moment Lagrangian. In Sect. \ref{Sec:CMDM-diagrams}, the six one-loop diagram contributions to the CMDM of the top quark are calculated. In Section \ref{Sec:results}, the numerical results for the CMDM of the top quark are discussed.
Sect. \ref{Sec:conclusions} is devoted to our conclusions.
Appendix \ref{appendix-Feynman-rules} presents the used input values.
Appendix \ref{appendix-PaVe} lists the resulting PaVes $A_0$, $B_0$ and $C_0$.
Appendix \ref{appendix-IR-regularization} demonstrates the DR of the IR divergent two-point PaVe from the CMDM no-Abelian diagram with the on-shell gluon. Appendix \ref{appendix-3g-parameterizations} addresses the IR divergence issue by both the FP method and the massive gluon artifice.

\begin{figure}[!t]
\begin{center}
\includegraphics[width=7.0cm]{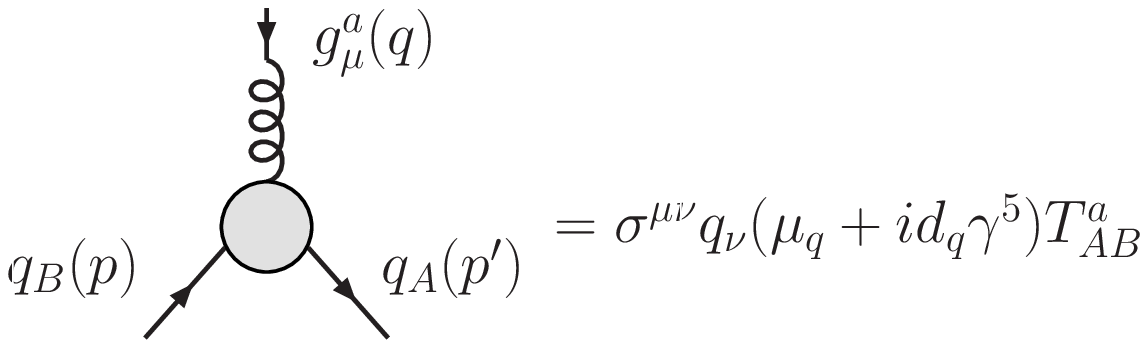}
\caption{Feynman rules for the chromoelectromagnetic dipole moments.}
\label{FIGURE-CEMDM-general}
\end{center}
\end{figure}

\section{The chromomagnetic dipole moment}
\label{Sec:CMDM-lagrangian}

The quark-antiquark-gluon interaction with the effective chromoelectromagnetic dipole moment (CEMDM) Lagrangian is \cite{Bernreuther:2013aga,Khachatryan:2016xws,Haberl:1995ek}
\begin{equation}\label{}
\mathcal{L}_{q\bar{q}g} =-g_s\thinspace\bar{q}_{A}\gamma^\mu q_{B}g_\mu^aT_{AB}^a+\mathcal{L}_\mathrm{eff},
\end{equation}
being
\begin{eqnarray}\label{lagrangian}
\mathcal{L}_\mathrm{eff} &=&-\frac{1}{2}\bar{q}_{A}\sigma^{\mu\nu}\left(\mu_{q}+id_{q}\gamma^5\right)q_{B}G_{\mu\nu}^aT_{AB}^a,
\end{eqnarray}
where $T_{AB}^a$ is the colour generator of $SU(3)_C$ ($A$ and $B$ are quark colour indices),
$\sigma^{\mu\nu}\equiv \frac{i}{2}[\gamma^\mu,\gamma^\nu]$,
$\mu_q$ is the CP-conserving chromomagnetic form factor, $d_q$ is the CP-violating chromoelectric (CEDM) form factor and
$G_{\mu\nu}^a=\partial_\mu g_\nu^a-\partial_\nu g_\mu^a-g_sf_{abc}g_\mu^bg_\nu^c$ is the gluon strength field, whose last term is not involved in the CMDM calculated below.
In the SM, the CMDM is induced perturbatively as a quantum fluctuation or radiative correction at the one-loop level \cite{Choudhury:2014lna,Bermudez:2017bpx,Martinez:2007qf}, while the CEDM arises at the three-loop level \cite{Czarnecki:1997bu}. Because the $\mathcal{L}_\mathrm{eff}$ have mass dimension 5, it is more suitable to define the dipoles as dimensionless \cite{PDG2020,Bernreuther:2013aga,Khachatryan:2016xws,Haberl:1995ek} as
\begin{equation}\label{}
\hat{\mu}_q\equiv \frac{m_q}{g_s}\mu_q, \quad \hat{d}_q\equiv \frac{m_q}{g_s}d_q,
\end{equation}
where $m_q$ is the quark mass, $g_s=\sqrt{4\pi\alpha_s}$ is the QCD group coupling constant, with $\alpha_s$ being the perturbative strong coupling constant, characterized at the mass of the $Z$ gauge boson, $\alpha_s(m_Z^2)=0.1179$ \cite{PDG2020}. In general, the CEMDM are complex quantities; they may have absorptive imaginary parts, for example, when the momentum transfer is timelike ($q^2>0$) in $t\bar{t}$ production via $pp$ collisions \cite{Bernreuther:2013aga,Khachatryan:2016xws}. Thus, from Eq. (\ref{lagrangian}), the CEMDM vertex or Feynman rule can be written as
\begin{equation}\label{}
\Gamma^\mu=\sigma^{\mu\nu}q_\nu\left(\mu_q+id_q\gamma^5\right)T_{AB}^a,
\end{equation}
where $q_\nu$ is the gluon momentum transfer, and $p+q=p'$ (see Fig.~\ref{FIGURE-CEMDM-general}).
The corresponding invariant amplitude is
\begin{equation}\label{}
\mathcal{M}=\mathcal{M}^\mu \epsilon^{a}_\mu(\vec{q})~,
\end{equation}
with the Lorentz structure
\begin{equation}\label{}
\mathcal{M}^\mu=\bar{u}(p')\Gamma^\mu u(p).
\end{equation}

\begin{figure*}[t!]
\begin{center}
\includegraphics[width=16.0cm]{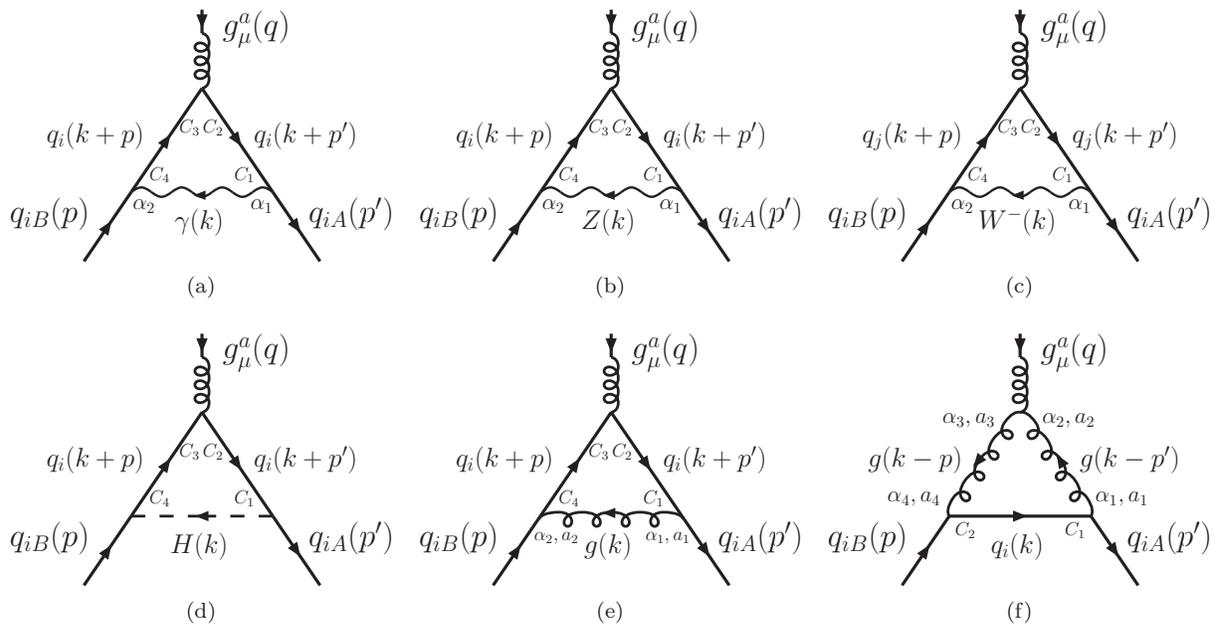}
\caption{CMDM in the SM: the EW contribution is the sum of the (a)-(d) diagrams, and the QCD contribution is the sum of (e) the Abelian and (f) the non-Abelian diagrams.}
\label{FIGURE-chromo}
\end{center}
\end{figure*}

\section{The CMDM in the SM at the one-loop level}
\label{Sec:CMDM-diagrams}

From now on, we will address the CMDM of the top quark, which we will refer to as $q_i=t$. The CMDM in the SM is composed of the sum of the six contributions
\begin{eqnarray}\label{CMDM-complete}
\hat{\mu}_{q_i}(q^2) &=& \hat{\mu}_{q_i}(\gamma)+\hat{\mu}_{q_i}(Z)+\hat{\mu}_{q_i}(W)+\hat{\mu}_{q_i}(H)
+\hat{\mu}_{q_i}(g)+\hat{\mu}_{q_i}(3g),
\end{eqnarray}
with each one depicted in Fig.~\ref{FIGURE-chromo}; we refer to them as
(a) $\hat{\mu}_{q_i}(\gamma)$ the Schwinger-type photon,
(b) $\hat{\mu}_{q_i}(Z)$ the $Z$ neutral gauge boson,
(c) $\hat{\mu}_{q_i}(W)$ the $W$ charged gauge boson,
(d) $\hat{\mu}_{q_i}(H)$ the Higgs boson,
(e) $\hat{\mu}_{q_i}(g)$ the Schwinger-type gluon,
and (f) $\hat{\mu}_{q_i}(3g)$ the triple gluon vertex.
The EW contribution comes from the sum of diagrams (a)-(d), and the QCD contribution comprises the sum of diagrams (e) and (f).

Below, general analytical results for the CMDM for each diagram with the off-shell gluon are first presented. Second, in case i), the on-shell gluon contributions ($q^2=0$) are computed to realize each expression but in particular to scrutinize the $\hat{\mu}_{q_i}(3g)$ diagram in order to appreciate its IR divergence details.
In case ii), the off-shell gluon ($q^2\neq 0$) evaluations are performed by plotting each CMDM as a function of the gluon momentum transfer $q^2=\pm M^2$, within the interval $M=[0,200]$ GeV; it is important to keep in mind that here, $M$ merely acts as a variable in GeV units and obviously must not be considered a mass of the external gluon.

The complete evaluation of the CMDM, coming from Eq.~(\ref{CMDM-complete}), will be addressed in Sect.~\ref{Sec:results}. In our analytical and numerical study, we have used the software
\texttt{Mathematica},
\texttt{FeynCalc} \cite{Mertig:1990an,Shtabovenko:2016sxi,Shtabovenko:2020gxv},
\texttt{FeynHelpers} \cite{Shtabovenko:2016whf},
\texttt{Package} \texttt{-X} \cite{Patel:2015tea},
\texttt{Rubi} \cite{Rubi},
\texttt{Collier} \cite{Denner:2016kdg} and LoopTools~\cite{Hahn:1998yk}.
Additionally, we have cross-checked our analytical calculations of the loop integrals versus those generated with \texttt{FeynArts} \cite{Hahn:2000kx}, as well as our numerical results for the CMDM contributions, by using the dedicated packages versus our own codes for the PaVes $A_0$~(\ref{A0-formula}), $B_0$~(\ref{B0-formula}) and $C_0$~(\ref{C0-formula}) given in Appendix~\ref{appendix-PaVe}.

\subsection{The $\gamma$ diagram}
\label{Sec:photon-diagram}

The Schwinger-type photon diagram is illustrated in Fig.~\ref{FIGURE-chromo}(a), where its tensor amplitude is
\begin{eqnarray}\label{photon-amplitude}
\mathcal{M}_{q_i}^\mu(\gamma) &=& \mu^{2\epsilon}\int\frac{d^Dk}{(2\pi)^D}\bar{u}(p')\left(-ieQ_{q_i}\gamma^{\alpha_1}\delta_{AC_1}\right)
\left[i\frac{\slashed{k}+\slashed{p}'+m_{q_i}}{(k+p')^2-m_{q_i}^2+i\varepsilon}\delta_{C_1C_2}\right]
\left(-ig_s\gamma^{\mu}T_{C_2C_3}^a\right)
\nonumber\\
&&\times
\left[i\frac{\slashed{k}+\slashed{p}+m_{q_i}}{(k+p)^2-m_{q_i}^2+i\varepsilon}\delta_{C_3C_4}\right]
\left(-ieQ_{q_i}\gamma^{\alpha_2}\delta_{C_4B}\right)
u(p)\left(i\frac{-g_{\alpha_1\alpha_2}}{k^2+i\varepsilon}\right),
\end{eqnarray}
with
$\delta_{AC_1}\delta_{C_1C_2}T_{C_2C_3}^a\delta_{C_3C_4}\delta_{C_4B}=T_{AB}^a$, where a sum over repeated indices is assumed.

After algebraic manipulations, the resulting part of the CMDM with the off-shell gluon ($q^2\neq0$) is
\begin{eqnarray}\label{MDM-photon}
\hat{\mu}_{q_i}(\gamma) &=& -\frac{\alpha Q_{q_i}^2m_{q_i}^2}{2\pi(q^2-4m_{q_i}^2)}
\left(B_{01}^{\gamma}-B_{02}^{\gamma}\right)
\nonumber\\
&=& \frac{\alpha Q_{q_i}^2 m_{q_i}^2 }{2 \pi  \sqrt{q^2 \left(q^2-4 m_{q_i}^2\right)}}
\ln\frac{\sqrt{q^2 \left(q^2-4 m_{q_i}^2\right)}+2 m_{q_i}^2-q^2}{2 m_{q_i}^2},
\end{eqnarray}
where $B_{01}^{\gamma}\equiv B_0\left(m_{q_i}^2,0,m_{q_i}^2\right)$ and
$B_{02}^{\gamma}\equiv B_0\left(q^2,m_{q_i}^2,m_{q_i}^2\right)$; the explicit form of the PaVes can be consulted in the Appendix~\ref{appendix-PaVe}.

i) On-shell gluon case ($q^2=0$): here, $\hat{\mu}_{q_i}(\gamma)$ is a constant,
with $B_{02}^{\gamma}\equiv B_0\left(0,m_{q_i}^2,m_{q_i}^2\right)$, and therefore,
\begin{equation}\label{}
\hat{\mu}_t(\gamma)=\frac{\alpha}{9\pi}.
\end{equation}
Analogous behaviour will occur for the Schwinger-type gluon diagram $\hat{\mu}_{q_i}(g)$ (see Fig.~\ref{FIGURE-chromo}(e)), as will be shown in Sect.~\ref{Sec:g-diagram}.

ii) Off-shell gluon case ($q^2=\pm m_Z^2$): from Eq. (\ref{MDM-photon}), it can be noticed that $\hat{\mu}_{q_i}(\gamma)\propto m_{q_i}^2$, which provides a large value for the top quark.

The resulting evaluations are listed in Table \ref{TABLE-chromo-top}, and the general behaviour of $\hat{\mu}_t(\gamma)$ is shown in Fig.~\ref{FIGURE-QCD-top}(a).

\subsection{The $Z$ gauge boson diagram}
\label{subsection-Z}

The $Z$ gauge boson contribution is shown in Fig.~\ref{FIGURE-chromo}(b), and the respective tensor amplitude is
\begin{eqnarray}
\mathcal{M}_{q_i}^\mu(Z) &=& \mu^{2\epsilon}\int\frac{d^Dk}{(2\pi)^D}\bar{u}(p')
\left[\frac{-ig}{2c_W}\gamma^{\alpha_1}(g_{Vq_i}-g_{Aq_i}\gamma^5)\delta_{AC_1}\right]
\left[i\frac{\slashed{k}+\slashed{p}'+m_{q_i}}{(k+p')^2-m_{q_i}^2+i\varepsilon}\delta_{C_1C_2}\right]
\left(-ig_s\gamma^{\mu}T_{C_2C_3}^a\right)
\nonumber\\
&& \times \left[i\frac{\slashed{k}+\slashed{p}+m_{q_i}}{(k+p)^2-m_{q_i}^2+i\varepsilon}\delta_{C_3C_4}\right]
\left[\frac{-ig}{2c_W}\gamma^{\alpha_2}\left(g_{Vq_i}-g_{Aq_i}\gamma^5\right)\delta_{C_4B}\right]
u(p)
\nonumber\\
&&\times\left[\frac{i}{k^2-m_{Z}^2+i\varepsilon}\left(-g_{\alpha_1\alpha_2}+\frac{k_{\alpha_1}k_{\alpha_2}}{m_{Z}^2}\right)\right],
\end{eqnarray}
with the same colour algebra as in Eq.~(\ref{photon-amplitude}).

The off-shell gluon ($q^2\neq 0$) CMDM induced by the $Z$ neutral gauge boson is
\begin{eqnarray}\label{MDM-Z}
\hat{\mu}_{q_i}(Z)&=& \frac{\alpha}{8\pi c_W^2s_W^2m_Z^2(q^2-4m_{q_i}^2)^2}
\Big(
g_{Vq_i}^2\big\{
m_Z^2\left(q^2-4m_{q_i}^2\right)\left(A_{01}^Z-A_{02}^Z+m_{q_i}^2\right)
\nonumber\\
&& +m_Z^2\left[8m_{q_i}^4-2m_{q_i}^2\left(5m_Z^2+q^2\right)+m_Z^2q^2\right]B_{01}^Z
+m_{q_i}^2m_Z^2\left(-4m_{q_i}^2+6m_Z^2+q^2\right)B_{02}^Z
\nonumber\\
&& +2m_{q_i}^2m_Z^4\left(-8m_{q_i}^2+3m_Z^2+2q^2\right)C_0^Z
\big\}
\nonumber\\
&& +g_{Aq_i}^2\big\{
\left(2m_{q_i}^2+m_Z^2\right)\left(q^2-4m_{q_i}^2\right)(A_{01}^Z-A_{02}^Z+m_{q_i}^2)
\nonumber\\
&& +m_Z^2\left[20m_{q_i}^4-2m_{q_i}^2\left(5m_Z^2+4q^2\right)+m_Z^2q^2\right]B_{01}^Z
+m_{q_i}^2\left[8 m_{q_i}^4-2m_{q_i}^2\left(12m_Z^2+q^2\right)+6m_Z^4+9m_Z^2q^2\right]B_{02}^Z
\nonumber\\
&&
+2m_{q_i}^2m_Z^2\left[24 m_{q_i}^4-2m_{q_i}^2\left(9m_Z^2+7q^2\right)+3m_Z^4+2(3m_Z^2+q^2)q^2\right]C_0^Z
\big\}
\Big),
\end{eqnarray}
with $A_{01}^Z$ $\equiv$ $A_0\left(m_{q_i}^2\right)$, $A_{02}^Z$ $\equiv$ $A_0\left(m_Z^2\right)$,
$B_{01}^Z$ $\equiv$ $B_0\big(m_{q_i}^2,m_{q_i}^2,$ $m_Z^2\big)$,
$B_{02}^Z$ $\equiv$ $B_0\left(q^2,m_{q_i}^2,m_{q_i}^2\right)$
and
$C_0^Z$  $\equiv$ $C_0\big(m_{q_i}^2,m_{q_i}^2,q^2,$ $m_{q_i}^2,m_Z^2,m_{q_i}^2\big)$; because the analytical formula for $C_0^Z$ with the off-shell gluon is very long, we present a suitable approximation (see Appendix~\ref{appendix-PaVe}).

i) On-shell gluon case ($q^2=0$): here, $B_{02}^Z\equiv B_0\big(0,m_{q_i}^2,$ $m_{q_i}^2\big)$
and $C_0^Z \equiv C_0\left(m_{q_i}^2,m_{q_i}^2,0,m_{q_i}^2,m_Z^2,m_{q_i}^2\right)$.

ii) Off-shell gluon case ($q^2=\pm m_Z^2$): its $C_0^Z$ function is given in Eq.~(\ref{C0-3}).

The numerical values of this contribution to the CMDM are given in Table~\ref{TABLE-chromo-top}, and its behaviour as a function of the gluon momentum transfer is displayed in Fig.~\ref{FIGURE-QCD-top}(b).

\subsection{The $W$ gauge boson diagram}
\label{subsection-W}

The $W$ gauge boson contribution (see Fig.~\ref{FIGURE-chromo}(c)) gives rise to the following tensor amplitude:
\begin{eqnarray}
\mathcal{M}_{q_i}^\mu(W) &=& \mu^{2\epsilon}\int\frac{d^Dk}{(2\pi)^D}\bar{u}(p')
\left(\frac{-ig}{\sqrt{2}}\gamma^{\alpha_1}P_LV_{q_iq_j}\delta_{AC_1}\right)
\left[i\frac{\slashed{k}+\slashed{p}'+m_{q_j}}{(k+p')^2-m_{q_j}^2+i\varepsilon}\delta_{C_1C_2}\right]
\left(-ig_s\gamma^{\mu}T_{C_2C_3}^a\right)
\nonumber\\
&&\times
\left[i\frac{\slashed{k}+\slashed{p}+m_{q_j}}{(k+p)^2-m_{q_j}^2+i\varepsilon}\delta_{C_3C_4}\right]
\left(\frac{-ig}{\sqrt{2}}\gamma^{\alpha_2}P_LV_{q_iq_j}^*\delta_{C_4B}\right)
u(p)
\nonumber\\
&&\times
\left[\frac{i}{k^2-m_W^2+i\varepsilon}\left(-g_{\alpha_1\alpha_2}+\frac{k_{\alpha_1}k_{\alpha_2}}{m_W^2}\right)\right],
\end{eqnarray}
being $q_1,q_2,q_3=d,s,b$, where the same colour algebra as in Eq.~(\ref{photon-amplitude}) applies.

The off-shell gluon ($q^2\neq 0$) CMDM induced by the $W$ gauge boson is composed by
\begin{eqnarray}
\hat{\mu}_{q_i}(W) &=& \sum_{j=1}^3\hat{\mu}_{q_i}(W,q_j)
\nonumber\\
&=& \frac{\alpha\sum_{j=1}^3|V_{q_iq_j}|^2}{16\pi s_W^2m_W^2 (q^2-4 m_{q_i}^2)^2}
\bigg[
\left(q^2-4 m_{q_i}^2\right)\left(m_{q_i}^2+m_{q_j}^2+2 m_W^2\right)\left(A_{01}^W-A_{02}^W+m_{q_i}^2\right)
\nonumber\\
&& +\left(-m_{q_i}^4\left[2 m_{q_i}^2+\left(8 m_{q_j}^2-18 m_W^2+q^2\right)\right]
+m_{q_i}^2\left\{2 m_{q_j}^2\left[5m_{q_j}^2+\left(5 m_W^2+q^2\right)\right]-m_W^2(20 m_W^2+9 q^2)\right\}
\right.
\nonumber\\
&& \left. -q^2 \left[m_{q_j}^4+m_W^2\left(m_{q_j}^2-2 m_W^2\right)\right]\right)B_{01}^W
\nonumber\\
&& +m_{q_i}^2 \left\{m_{q_i}^2\left[2 m_{q_i}^2+\left(12 m_{q_j}^2-22 m_W^2+q^2\right)\right]
-m_{q_j}^2\left[6 m_{q_j}^2+3\left(2 m_W^2+q^2\right)\right]+2 m_W^2 \left(6 m_W^2+5 q^2\right)\right\}B_{02}^W
\nonumber\\
&& -2 m_{q_i}^2 \left(m_{q_i}^4\left[m_{q_i}^2-\left(5 m_{q_j}^2+12 m_W^2+q^2\right)\right]
+m_{q_i}^2\left\{m_{q_j}^2\left[7m_{q_j}^2+2\left(q^2-6 m_W^2\right)\right]+m_W^2\left(17 m_W^2+8 q^2\right)\right\}
\right.
\nonumber\\
&& \left. -m_{q_j}^2\left[m_{q_j}^2(3 m_{q_j}^2+q^2)-m_W^2\left(9 m_W^2+6q^2\right)\right]
-2 m_W^2 \left(m_W^2+q^2\right) \left(3 m_W^2+q^2\right)\right)C_0^W \bigg],
\end{eqnarray}
where
$A_{01}^W$ $\equiv$ $A_0(m_{q_j}^2)$,
$A_{02}^W$ $\equiv$ $A_0(m_W^2)$,
$B_{01}^W$ $\equiv$ $B_0(m_{q_i}^2, m_{q_j}^2,$ $m_W^2)$,
$B_{02}^W$ $\equiv$ $B_0(q^2, m_{q_j}^2, m_{q_j}^2)$, and
$C_0^W$ $\equiv$ $C_0(m_{q_i}^2, m_{q_i}^2, q^2, m_{q_j}^2,$ $m_W^2, m_{q_j}^2)$.

i) On-shell gluon case ($q^2=0$): here,
$B_{02}^W$$\equiv$$(0,m_{q_j}^2,m_{q_j}^2)$
and
$C_0^W$$\equiv$$C_0(m_{q_i}^2, m_{q_i}^2, 0, m_{q_j}^2, m_W^2, m_{q_j}^2)$.

ii) Off-shell gluon case ($q^2=\pm m_Z^2$): since the $C_0^W$ solution is extremely long, instead, we provide the FP formula for its numerical evaluation (see Eq.~(\ref{C0-formula})).

Table~\ref{TABLE-chromo-top} lists the resulting values for the CMDM, where it should be noted that it has real and imaginary parts. Fig.~\ref{FIGURE-QCD-top}~(c) shows the real part, and Fig.~\ref{FIGURE-QCD-top}~(d) shows the imaginary one.

\subsection{The Higgs boson diagram}
\label{subsection-H}

The Higgs boson contribution (see Fig.~\ref{FIGURE-chromo}(d)) generates the tensor amplitude
\begin{eqnarray}
\mathcal{M}_{q_i}^\mu(H) &=& \mu^{2\epsilon}\int\frac{d^Dk}{(2\pi)^D}\bar{u}(p')
\left(\frac{-igm_{q_i}}{2m_W}\delta_{AC_1}\right)
\left[i\frac{\slashed{k}+\slashed{p}'+m_{q_i}}{(k+p')^2-m_{q_i}^2+i\varepsilon}\delta_{C_1C_2}\right]
\left(-ig_s\gamma^{\mu}T_{C_2C_3}^a\right)
\nonumber\\
&&\times\left[i\frac{\slashed{k}+\slashed{p}+m_{q_i}}{(k+p)^2-m_{q_i}^2+i\varepsilon}\delta_{C_3C_4}\right]
\left(\frac{-igm_{q_i}}{2m_W}\delta_{C_4B}\right)
u(p) \left(\frac{i}{k^2-m_H^2+i\varepsilon}\right).
\end{eqnarray}
Again, the colour algebra is the same as in Eq.~(\ref{photon-amplitude}).

The off-shell gluon ($q^2\neq 0$) CMDM generated by the Higgs boson is
\begin{eqnarray}
\hat{\mu}_{q_i}(H) &=& \frac{\alpha~m_{q_i}^2}{16\pi m_W^2s_W^2\left(q^2-4m_{q_i}^2\right)^2}
\big\{\left(q^2-4m_{q_i}^2\right)\left(A_{01}^H-A_{02}^H+m_{q_i}^2\right)
\nonumber\\
&& +\left[16m_{q_i}^4-2m_{q_i}^2\left(5m_H^2+2q^2\right)+m_H^2q^2\right]B_{01}^H
+3m_{q_i}^2\left(-4m_{q_i}^2+2m_H^2+q^2\right)B_{02}^H
\nonumber\\
&& +6m_{q_i}^2m_H^2\left(-4 m_{q_i}^2+m_H^2+q^2\right)C_0^H
\big\},
\end{eqnarray}
with
$A_{01}^H$ $\equiv$ $A_0(m_{q_i}^2)$,
$A_{02}^H$ $\equiv$ $A_0(m_H^2)$,
$B_{01}^H$ $\equiv$ $B_0(m_{q_i}^2, m_{q_i}^2,$ $m_H^2)$,
$B_{02}^H$ $\equiv$ $B_0(q^2, m_{q_i}^2, m_{q_i}^2)$,
and $C_0^H\equiv C_0(m_{q_i}^2, m_{q_i}^2, q^2,$ $m_{q_i}^2, m_H^2, m_{q_i}^2)$.

i) On-shell gluon case ($q^2=0)$: for $B_{02}^H$ $\equiv$ $B_0\big(0,m_{q_i}^2,$ $m_{q_i}^2\big)$
and $C_0^H$ $\equiv$ $C_0\left(m_{q_i}^2,m_{q_i}^2,0,m_{q_i}^2,m_H^2,m_{q_i}^2\right)$.

ii) Off-shell gluon case ($q^2=\pm m_Z^2$): the $C_0^H$ analytical approximation for this situation is given in Eq. (\ref{C0-3}).

All the resulting values are listed in Table \ref{TABLE-chromo-top}. The plot of this contribution can be seen in Fig.~\ref{FIGURE-QCD-top}(e).

\subsection{The $g$ diagram}
\label{Sec:g-diagram}

The Schwinger-type gluon diagram (see Fig.~\ref{FIGURE-chromo}(e)) offers this tensor amplitude
\begin{eqnarray}
\mathcal{M}_{q_i}^\mu(g) &=& \mu^{2\epsilon}\int\frac{d^Dk}{(2\pi)^D}\bar{u}(p')\left(-ig_s\gamma^{\alpha_1}T_{AC_1}^{a_1}\right)
\left[i\frac{\slashed{k}+\slashed{p}'+m_{q_i}}{(k+p')^2-m_{q_i}^2+i\varepsilon}\delta_{C_1C_2}\right]
\left(-ig_s\gamma^{\mu}T_{C_2C_3}^a\right)
\nonumber\\
&&\times
\left[i\frac{\slashed{k}+\slashed{p}+m_{q_i}}{(k+p)^2-m_{q_i}^2+i\varepsilon}\delta_{C_3C_4}\right]
\left(-ig_s\gamma^{\alpha_2}T_{C_4B}^{a_2}\right)
u(p)
\left(i\frac{-g_{\alpha_1\alpha_2}}{k^2+i\varepsilon}\delta_{a_1a_2}\right),
\end{eqnarray}
where
$T_{AC_1}^{a_1}\delta_{C_1C2}T_{C_2C_3}^a\delta_{C_3C_4}T_{C_4B}^{a_2}\delta_{a_1a_2}$ $=$
$T_{AC_2}^{a_1}T_{C_2C_4}^aT_{C_4B}^{a_1}$ $=$
$(T^{a_1}T^aT^{a_1})_{AB}$ $=$
$(C_F-\frac{1}{2}C_A)T_{AB}^a=-\frac{1}{6}T_{AB}^a$, $C_A=N=3$, $C_F$=$(N^2-1)/2N$=$4/3$.

The resulting off-shell gluon CMDM is
\begin{eqnarray}\label{CMDM-gluon}
\hat{\mu}_{q_i}(g) &=& \frac{\alpha_sm_{q_i}^2}{12\pi(q^2-4m_{q_i}^2)}
\left(B_{01}^g-B_{02}^g\right)
\nonumber\\
&=& -\frac{\alpha_s m_{q_i}^2 }{12 \pi  \sqrt{q^2 \left(q^2-4 m_{q_i}^2\right)}}
\ln\frac{\sqrt{q^2 \left(q^2-4 m_{q_i}^2\right)}+2 m_{q_i}^2-q^2}{2 m_{q_i}^2}~,
\end{eqnarray}
being $B_{01}^g\equiv B_0\left(m_{q_i}^2,0,m_{q_i}^2\right)$,
$B_{02}^g\equiv B_0\left(q^2,m_{q_i}^2,m_{q_i}^2\right)$. As already commented, this virtual gluon case is entirely analogous to the Schwinger-type photon case from Sect. \ref{Sec:photon-diagram}.

i) On-shell gluon case ($q^2=0$): here, the result is
\begin{equation}\label{}
\hat{\mu}_{q_i}(g)=-\frac{\alpha_s}{24\pi}.
\end{equation}

ii) Off-shell gluon case ($q^2=\pm m_Z^2$): from Eq.~(\ref{CMDM-gluon}), it is clear that $\hat{\mu}_{q_i}(g)\propto m_{q_i}^2$, which yields a large contribution for the top quark CMDM.

The respective numerical evaluations are listed in Table~\ref{TABLE-chromo-top}, and the corresponding CMDM behaviour is presented in Fig.~\ref{FIGURE-QCD-top}(f).

\subsection{The $3g$ diagram}
\label{subsection-3g}

The triple gluon vertex diagram, characterized by being the only non-Abelian contribution to the CMDM, contains an IR divergence when the gluon is on-shell that the previous literature has not properly addressed, which is why we treated it in detail. The associated Feynman diagram is depicted in Fig.~\ref{FIGURE-chromo}(f), for which the tensor amplitude is written as
\begin{eqnarray}\label{3g-integral}
\mathcal{M}_{q_i}^\mu(3g) &=& \mu^{2\epsilon}\int\frac{d^Dk}{(2\pi)^D}\bar{u}(p')\left(-ig_s\gamma^{\alpha_1}T_{AC_1}^{a_1}\right)
\left(i\frac{\slashed{k}+m_{q_i}}{k^2-m_{q_i}^2+i\varepsilon}\delta_{C_1C_2}\right)\left(-ig_s\gamma^{\alpha_4}T_{C_2B}^{a_4}\right)u(p)
\nonumber\\
&&\times
\left[\frac{i}{(k-p')^2+i\varepsilon}
\left(-g_{\alpha_1\alpha_2}\right)
\delta_{a_1a_2}\right]
\left[-g_sf_{aa_3a_2}T^{\mu\alpha_3\alpha_2}_{ggg}(p'-p,-k+p,k-p')\right]
\nonumber\\
&&\times
\left[\frac{i}{(k-p)^2+i\varepsilon}
\left(-g_{\alpha_3\alpha_4}\right)
\delta_{a_3a_4}\right]
\nonumber\\
&=&-\frac{3g_s^3}{2}T_{AB}^a
\mu^{2\epsilon}\int\frac{d^Dk}{(2\pi)^D}
\frac{\bar{u}(p')\gamma^{\alpha_1}(\slashed{k}+m_{q_i})\gamma^{\alpha_4}u(p)}
{\left(k^2-m_{q_i}^2+i\varepsilon\right)\left[(k-p')^2+i\varepsilon\right]\left[(k-p)^2+i\varepsilon\right]}
\nonumber\\
&&\times
\left(-g_{\alpha_1\alpha_2}\right)
T^{\mu\alpha_3\alpha_2}_{ggg}(p'-p,-k+p,k-p')
\left(-g_{\alpha_3\alpha_4}\right),
\end{eqnarray}
with the colour algebra\footnote{It is important to highlight that in Eq.~(9) from Ref. \cite{Choudhury:2014lna}, their  $T_{ji'}^cT_{i'i}^bf_{abc}=-iT_{ji}^a/4$ is incorrect.}
$T_{AC_1}^{a_1}\delta_{C_1C_2}T_{C_2B}^{a_4}\delta_{a_1a_2}f_{aa_3a_2}\delta_{a_3a_4}$ $=$
$(-f_{aa_2a_3}T^{a_2}T^{a_3})_{AB}$ $=$
$-\frac{i}{2}C_AT_{AB}^a$ $=$
$-i\frac{3}{2}T_{AB}^a$, and the tensor
$T_{\mu\alpha_3\alpha_2}^{ggg}(p'-p,-k+p,k-p')$ $\equiv$
$(k-2p+p')_{\alpha_2}g_{\mu\alpha_3}+(-2k+p+p')_\mu g_{\alpha_3\alpha_2}+(k+p-2p')_{\alpha_3}g_{\alpha_2\mu}$.

After applying DR to the above amplitude, the CMDM with the off-shell gluon ($q^2\neq 0$) can be extracted, being
\begin{eqnarray}\label{MDM-3g}
\hat{\mu}_{q_i}(3g) &=&
\frac{3\alpha_s}{4\pi} \frac{m_{q_i}^4}{(q^2-4m_{q_i}^2)^2}
\left[8-\frac{2 q^2}{m_{q_i}^2}+\left(8+\frac{q^2}{m_{q_i}^2}\right)
\right.
\left.
\big(B_{01}^{3g}-B_{02}^{3g}\big)-6q^2C_0^{3g}
\right],
\end{eqnarray}
with $B_{01}^{3g} \equiv B_0(m_{q_i}^2,0,m_{q_i}^2)$,
$B_{02}^{3g} \equiv B_0(q^2,0,0)$, and
$C_0^{3g} \equiv C_0(m_{q_i}^2, m_{q_i}^2, q^2, 0, m_{q_i}^2, 0)$ (see Appendix~\ref{appendix-PaVe}). We emphasize that this contribution of the CMDM is strictly valid only when $q^2\neq 0$.

Nonetheless, an IR divergence arises when $q^2\to 0$: specifically, from the part (in Eq.~(\ref{MDM-3g}))
\begin{eqnarray}\label{IR-divergence-1}
B_{01}^{3g}-B_{02}^{3g} = -\ln\frac{m_{q_i}^2}{-q^2}.
\end{eqnarray}
This behaviour comes from $B_{02}^{3g}$; for more details, see Eq.~(\ref{B0-3g}) in Appendix \ref{appendix-PaVe}.
Then, by considering sufficiently small $q^2$, we obtain
\begin{eqnarray}\label{IR-limit}
\hat{\mu}_{q_i}(3g) \approx \frac{3\alpha_s}{8\pi}
\left(1-\ln\frac{~m_{q_i}^2}{-q^2}\right),
\end{eqnarray}
which diverges if $q^2\to 0$.

This problematic logarithm in Eq. (\ref{IR-limit}) was also pointed out in Eq.~(37) from Ref.~\cite{Bermudez:2017bpx}, but the source that induces the IR divergence was not indicated. On the other hand, in Eq.~(11) from Ref.~\cite{Choudhury:2014lna}, the IR divergence was presented through the FP method without considering the $+i\varepsilon$ prescription.

\begin{figure*}[t!]
\begin{center}
\includegraphics[scale=1]{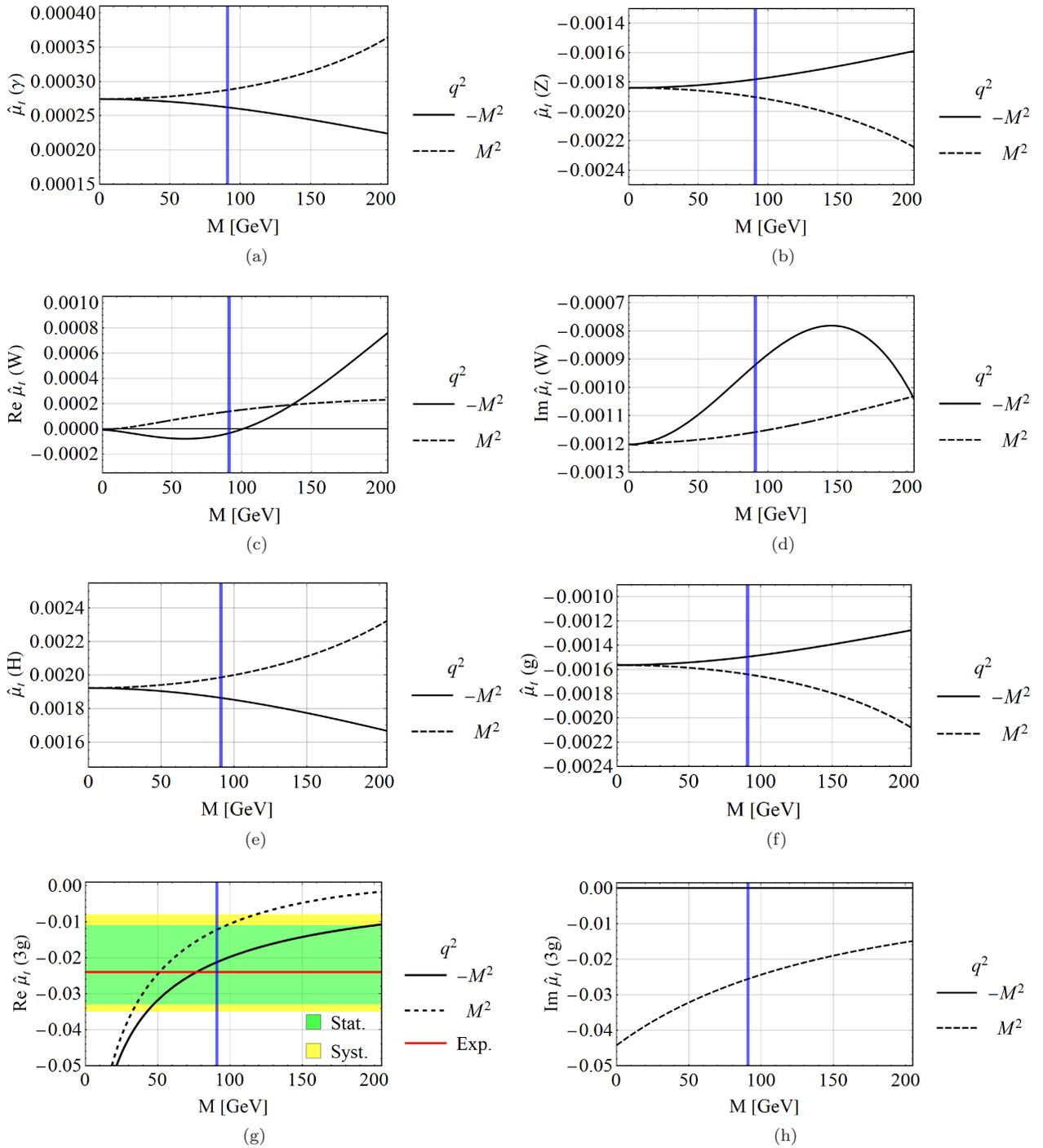}
\caption{Contributions to the CMDM of the top quark as function of the gluon momentum transfer $q^2=\pm M^2$, where $M=[0,200]$ GeV; the blue vertical line indicates $M=m_Z$. The largest contribution to the CMDM of the top quark, $\mathrm{Re}\thinspace\hat{\mu}_t(3g)$, is shown in (g), where it is compared with the experimental measure $\hat{\mu}_t^\mathrm{Exp}$.}
\label{FIGURE-QCD-top}
\end{center}
\end{figure*}

To delve into details of the IR singularity and provide a wide panorama of the different approaches for dealing with it, we present four different schemes that lead to the same divergent issue. This can be appreciated in Appendices \ref{appendix-IR-regularization} and \ref{appendix-3g-parameterizations}. First, in Appendix~\ref{appendix-IR-regularization}, we treat the IR problem by using DR, which represents the most formal procedure in quantum field theory. Second, in Appendix \ref{appendix-3g-parameterizations}, we focus on the problem by applying the FP method considering the $+i\varepsilon$ Feynman prescription in all the propagators, which is crucial to keep track of the IR divergence in the triple gluon vertex contribution. We begin employing the gluon propagator in the Feynman-'t Hooft gauge $\xi=1$; afterwards, the general renormalizable $R_\xi$ gauge is taken into account. Finally, we also apply the fictitious mass regularization scheme for virtual gluons or massive gluons artifice. In summary, all of these different procedures reveal the same IR divergence issue when the gluon is on-shell.

i) On-shell gluon case ($q^2\to 0$): the resulting dimensional regularized two-point scalar function $B_{02}^{3g} \equiv B_0(q^2,0,0)$, for $q^2\to0$, is given in Eq.~(\ref{B0-3g-final}), which is now expressed in terms of the UV and IR poles as
\begin{eqnarray}
B_{02}^{3g} &\equiv& B_0(0,0,0)
\nonumber\\
&=&
\frac{1}{\epsilon_\mathrm{UV}}-\frac{1}{\epsilon_\mathrm{IR}}
\nonumber\\
&=&
\Delta_\mathrm{UV}-\Delta_\mathrm{IR},
\end{eqnarray}
with $\Delta_\mathrm{UV}$ and $\Delta_\mathrm{IR}$ defined in Eqs. (\ref{DivergenceUV}) and (\ref{DivergenceIR}), respectively. In addition, the last term from Eq. (\ref{MDM-3g}) vanishes
\begin{equation}
q^2C_0^{3g}=0,
\end{equation}
when $q^2\to 0$. Therefore, the CMDM from the triple gluon vertex diagram given in Eq.~(\ref{MDM-3g}), with the on-shell gluon, takes the final form
\begin{equation}\label{MDM-3g-on-shell}
\lim_{q^2\to 0}\hat{\mu}_{q_i}(3g) =
\frac{3\alpha_s}{8\pi} \left(\Delta_\mathrm{IR}+\ln\frac{\mu^2}{m_{q_i}^2}+3\right),
\end{equation}
where $\Delta_\mathrm{IR}$ contains the pole $1/\epsilon_\mathrm{IR}$ of IR nature.
Numerically, this divergent behaviour can also be appreciated, for the top quark, in Fig. \ref{FIGURE-QCD-top}~(g) when $q^2=\pm M\to 0$.

ii) Off-shell gluon case ($q^2=\pm m_Z^2$): from Eq.~(\ref{MDM-3g}), the spacelike value, $q^2=-m_Z^2$, only yields a real part, while the timelike value, $q^2=m_Z^2$, provides real and imaginary parts; these values are listed in Table~\ref{TABLE-chromo-top}. The behaviour of $\hat{\mu}_t(3g)$ as a function of $q^2=\pm M^2$ is shown in Figs.~\ref{FIGURE-QCD-top}(g) and (h).

\begin{table*}[!t]
  \centering
\begin{tabular}{|c|c|c|c|}\hline
\multirow{2}{*}{$\hat{\mu}_t$} & \multicolumn{3}{c|}{$q^2$} \\
  \cline{2-4}
         & $-m_Z^2$              & 0                      & $m_Z^2$  \\
\hline
$\gamma$ & $2.62\times10^{-4}$  & $2.74\times10^{-4}$   & $2.88\times10^{-4}$  \\
$Z$      & $-1.78\times10^{-3}$ & $-1.84\times10^{-3}$  & $-1.90\times10^{-3}$ \\
$W$      & $-2.91\times10^{-5}-9.25\times10^{-4}i$      & $6.29\times10^{-7}-1.21\times10^{-3}i$ & $1.44\times10^{-4}-1.16\times10^{-3}i$ \\
$H$      & $1.86\times10^{-3}$  & $1.92\times10^{-3}$   & $1.99\times10^{-3}$  \\
$g$      & $-1.50\times10^{-3}$ & $-1.56\times10^{-3}$  & $-1.64\times10^{-3}$ \\
$3g$     & $-2.12\times10^{-2}$ & $\mathrm{IR~div.}$    & $-1.22\times10^{-2}-2.55\times10^{-2}i$ \\
Total    & $-2.24\times10^{-2}-9.25\times10^{-4}i$  & $\mathrm{IR~div.}$ & $-1.33\times10^{-2}-2.67\times10^{-2}i$ \\
\hline
\end{tabular}
\caption{The different contributions to the top quark CMDM; the experimental value is $\hat{\mu}_t^\mathrm{Exp}$ $=$ $-0.024_{-0.009}^{+0.013}(\mathrm{stat})_{-0.011}^{+0.016}(\mathrm{syst})$ \cite{Sirunyan:2019eyu}. }\label{TABLE-chromo-top}
\end{table*}

\begin{table*}[!t]
  \centering
\begin{tabular}{|c|c|c|c|}\hline
\multirow{2}{*}{$\hat{\mu}_t$} & \multicolumn{3}{c|}{$q^2$} \\
  \cline{2-4}
         & $-m_Z^2$               & 0                      & $m_Z^2$  \\
\hline
EW       & $0.000315-0.000925i$ & $0.000357-0.00121i$  & $0.000514-0.00116i$    \\
QCD      & $-0.0227$ & $\mathrm{IR~div.}$ & $-0.0138-0.0255i$    \\
Total    & $-0.0224-0.000925i$ & $\mathrm{IR~div.}$    & $-0.0133-0.0267i$    \\
\hline
\end{tabular}
\caption{The top quark CMDM separated into EW and QCD contributions; the experimental value is $\hat{\mu}_t^\mathrm{Exp}=-0.024_{-0.009}^{+0.013}(\mathrm{stat})_{-0.011}^{+0.016}(\mathrm{syst})$ \cite{Sirunyan:2019eyu}. }\label{TABLE-chromo-top-parts}
\end{table*}

\section{Results}
\label{Sec:results}

It is important to note that the CMDM of the top quark in the SM is gauge-independent even though $q^2\neq0$. Our calculations have been tested for the unitary, Feynman-'t Hooft and $R_\xi$ gauges, resulting in the same predictions for $q^2\neq0$. Moreover, in this work, we prove that the triple gluon contribution at the one-loop level is also gauge-independent when $q^2\to0$, where the corresponding amplitude is considered in the context of the general $R_\xi$ gauge, which can be appreciated in detail in Appendix \ref{appendix-3g-FP-Rxi}. In this regard, we have corroborated the gauge-independence even when $q^2\neq0$, by computing the one-loop integral in Eq.~(\ref{3g-integral-Rxi}) by means of the Passarino--Veltman tensor decomposition method, which agrees with the final result shown in Eq.~(\ref{MDM-3g}).

As already mentioned in the Introduction and in Sect. \ref{subsection-3g}, because of the IR divergence presence in $\hat{\mu}_{q_i}(3g)$ when $q^2\to 0$, the complete $\hat{\mu}_{q_i}$ in Eq.~(\ref{CMDM-complete}) cannot be evaluated with the on-shell gluon; nevertheless, as proposed in Ref.~\cite{Choudhury:2014lna}, it makes sense to evaluate it at a conventional large gluon momentum transfer scale $q^2=\pm m_Z^2$, as is the case for the perturbative strong running coupling constant $\alpha_s(-q^2=m_Z^2)=0.1179$, which is conceived in the spacelike domain $q^2<0$ \cite{Field:1989uq,Deur:2016tte,BeiglboCk:2006lfa,Nesterenko:2016pmx}. Regarding the numerical results reported in the present work, we use updated input values taken from the PDG 2020 \cite{PDG2020} (see Appendix~\ref{appendix-Feynman-rules}).

The $\hat{\mu}_t$ evaluations for $q^2=-m_Z^2,0,m_Z^2$ for each contribution are listed in Table~\ref{TABLE-chromo-top}. Figure \ref{FIGURE-QCD-top} shows the six different contributions to the top quark CMDM by displaying the behaviour of each individual CMDM as a function of the gluon momentum transfer $q^2=\pm M^2$; the value $M=m_Z$ of the interval $M=[0,200]$ GeV is highlighted with the vertical blue line. The on-shell gluon values are included, which are finite for all contributions except for the triple gluon vertex. In Table~\ref{TABLE-chromo-top}, it can be appreciated that the on-shell gluon $q^2=0$ evaluation for $\mathrm{Re}\,\hat{\mu}_{t}(X)$ (for $X=\gamma, Z, W, H, g$) essentially corresponds to the central value of the two other values at $q^2=\pm m_Z^2$; it is remarkable that these resulting values are very close to each other. Figure \ref{FIGURE-QCD-top}(a) presents the $\hat{\mu}_t(\gamma)$ Schwinger-type photon contribution; the values are entirely real and positive. Fig.~\ref{FIGURE-QCD-top}(b) displays the behaviour of the $Z$ neutral gauge boson contribution $\hat{\mu}_t(Z)$; this provides only negative real values that are one order of magnitude larger and even closer to each other than in the photon case. Figure \ref{FIGURE-QCD-top}~(c) and (d) correspond to the $W$ charged gauge boson contribution $\hat{\mu}_t(W)$, which includes real and imaginary parts: the $\mathrm{Re}\thinspace\hat{\mu}_t(W)$ part is plotted in (c), and its $\mathrm{Im}\thinspace\hat{\mu}_t(W)$ part is plotted in (d). Notice that at $q^2=\pm m_Z^2$, the magnitude of the imaginary part is one order of magnitude larger than the real part, and these values are mainly due to the virtual bottom quark.
Figure \ref{FIGURE-QCD-top}(e) exhibits the Higgs boson contribution $\hat{\mu}_t(H)$, where it should be noted that the values are quite similar to those of the $Z$ case, but with opposite sign.
Figure \ref{FIGURE-QCD-top}(f) presents the Schwinger-type gluon contribution $\hat{\mu}_t(g)$, which only has a real part that is of the same order of magnitude $\sim10^{-3}$ as the $Z$ and $H$ cases.

\begin{figure*}[!t]
\begin{center}
\includegraphics[scale=1]{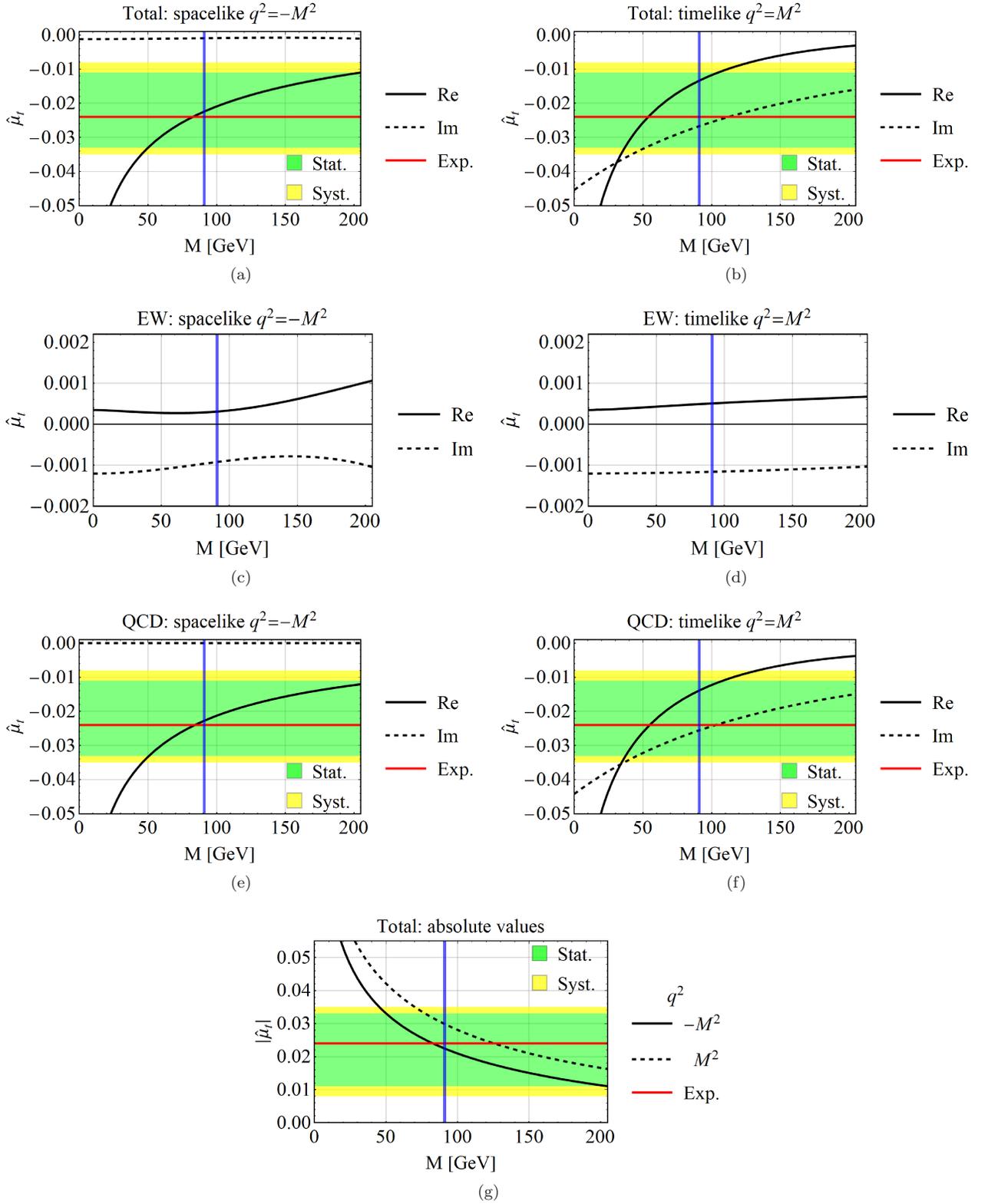}
\caption{Top quark CMDM as a function of the gluon momentum transfer $q^2=\pm M^2$, where $M=[0,200]$ GeV; the blue line indicates $M=m_Z$ and the experimental value $\hat{\mu}_t^\mathrm{Exp}$ is displayed. In (a) and (b), the total contributions are shown, in (c) and (d), the EW parts are depicted, and in (e) and (f), the QCD parts are presented. In (g), the absolute values of the total contributions for the spacelike domain, the timelike domain and the experimental value are compared.}
\label{FIGURE-QCD-top-2}
\end{center}
\end{figure*}

The triple gluon vertex contribution, $\hat{\mu}_t(3g)$, is shown in Fig. \ref{FIGURE-QCD-top}(g) and (h), where its real and imaginary parts are displayed, respectively. Because $\hat{\mu}_t(3g)$ is responsible for the largest value to the complete CMDM, $\hat{\mu}_t$, at $q^2=\pm m_Z^2$ (see Table~\ref{TABLE-chromo-top}), we compare it with the central experimental measure given in Eq.~(\ref{CMDM-experiment}), which we indicate in the plots with a horizontal red line. In Fig. \ref{FIGURE-QCD-top}(g), the $\mathrm{Re}\thinspace\hat{\mu}_t(3g)$ part with $q^2=\pm M^2$ is plotted, and both curves manifest the IR divergence feature when the external gluon is on-shell; notice that the spacelike evaluation produces only real values, and its imaginary part is exactly zero. On the other hand, the timelike evaluation yields a complex quantity, which can be appreciated in Fig. \ref{FIGURE-QCD-top}(h).

The total CMDM of the top quark $\hat{\mu}_t(q^2)$ is shown in Fig. \ref{FIGURE-QCD-top-2}(a) and (b) for the spacelike and the timelike domains, respectively. Our main result is the spacelike evaluation (see Table \ref{TABLE-chromo-top-parts}), since the $\mathrm{Re}\thinspace\hat{\mu}_t(-m_Z^2)$ part matches with the experimental central value $\hat{\mu}_t^\mathrm{Exp}=-0.024_{-0.009}^{+0.013}(\mathrm{stat})_{-0.011}^{+0.016}(\mathrm{syst})$. In contrast, our result for the $\mathrm{Im}\thinspace\hat{\mu}_t(-m_Z^2)$ part is purely an EW effect induced by the $W$ gauge boson loop. Moreover, the timelike domain produces also complex values (see Fig.~\ref{FIGURE-QCD-top-2}(b)), and its real part is entirely within the experimental statistical error; here, the $\hat{\mu}_t(3g)$ generates by itself an imaginary part.

The EW contributions are displayed in Fig. \ref{FIGURE-QCD-top-2}(c) and (d), which are given by $\hat{\mu}_t(q^2)^\mathrm{EW}$$=$$\hat{\mu}_{t}(\gamma)$ $+$ $\hat{\mu}_{t}(Z)$ $+$ $\hat{\mu}_{t}(W)$ $+$ $\hat{\mu}_{t}(H)$ (see Table~\ref{TABLE-chromo-top-parts}). The QCD contributions are displayed in Fig.~\ref{FIGURE-QCD-top-2}(e) and (f), and these are composed by $\hat{\mu}_t(q^2)^\mathrm{QCD}$ $=$ $\hat{\mu}_t(g)$ $+$ $\hat{\mu}_t(3g)$ (see Table~\ref{TABLE-chromo-top-parts}). It is obvious that both Fig. \ref{FIGURE-QCD-top-2}(e) and (f) essentially resemble the respective (a) and (b) plots, which is because the EW contribution is up to two orders of magnitude smaller than the QCD contribution.

It is worth comparing the absolute values of our results with the experimental values, as plotted in Fig. \ref{FIGURE-QCD-top-2}(g), where
$\big|\hat{\mu}_t(-m_Z^2)\big|$$=$ $0.0224$,
$\big|\hat{\mu}_t(m_Z^2)\big|$$=$ $0.0298$ and
$\big|\hat{\mu}_t^\mathrm{Exp}\big|$$=$ $0.024$.

By comparing our results with previous findings in the literature, we realized that Ref.~\cite{Choudhury:2014lna} is the only study that has evaluated the CMDM of the top quark at the $m_Z$ scale in the SM. Specifically, this work only considers the spacelike case, $q^2=-m_Z^2$, finding predictions for the form factor $\Delta\kappa$. In contrast, we compute the conventional $\hat{\mu}$, used in theoretical and experimental studies~\cite{Sirunyan:2019eyu,Bernreuther:2013aga,Khachatryan:2016xws,Haberl:1995ek}, with equivalence as $\hat{\mu}=\Delta\kappa/4$. Taking this consideration into account, from Table~2 \cite{Choudhury:2014lna}, devoted to the EW contributions, the  following is found: $\hat{\mu}_t(\gamma)$ $=$ $2.64\times10^{-4}$, $\hat{\mu}_t(Z)$ $=$ $6.42\times10^{-4}$, $\hat{\mu}_t(W)$ $=$ $-7.18\times10^{-5}$, $\hat{\mu}_t(H)$ $=$ $3.75\times10^{-3}$. On the other hand, our predictions only agree with their photon case ($\hat{\mu}_t(\gamma)$); for the Higgs contribution, our result is of the same order of magnitude; for the $Z$ contribution, our estimation is one order of magnitude larger; and for the $W$ boson contribution, our result for the real part of $\hat{\mu}_t(W)$ is of the same order of magnitude. However, they do not report the imaginary part. Notice that we have cross-checked all our results in three different ways: using the two different tools \texttt{Package-X} and \texttt{Collier}, and by our own PaVe formulas given in Eqs.~(\ref{A0-formula}), (\ref{B0-formula}) and (\ref{C0-formula}). Moreover, the authors did not provide explicit individual evaluations for each QCD diagram; that is, the Schwinger-type gluon diagram and the triple gluon vertex diagram. In Fig.~5(e), they show the total value for the CMDM of the top quark, being $\Delta\kappa_t=0.0264786$ and corresponding to $\hat{\mu}_t=0.00662$, which is one order of magnitude lower than our total prediction (see also footnote 1). Nevertheless, the authors of this reference realize that the triple gluon vertex contribution contains an infrared divergence, which is expressed as a logarithmic gluon mass singularity.

Reference \cite{Bermudez:2017bpx} presents the CMDM infrared divergence for the triple gluon vertex contribution for any quark of the SM. In particular, this singularity is established in logarithmic form in terms of the gluon momentum transfer, such as that shown in our Eq.~(\ref{IR-limit}). Nonetheless, this work does not offer phenomenological evaluations for the SM top quark.

In a previous work~\cite{Aranda:2018zis}, the CMDM of the SM top quark was numerically evaluated, finding very similar results to those presented in this new paper. The previous work used input values corresponding to the PDG 2018 version; in this paper, essentially the same results are found by considering current input values (PDG 2020)~\cite{PDG2020}. In addition, an imaginary contribution for $\hat{\mu}_t(W)$ was reported. In contrast, in this work we offer detailed analytical calculations for each diagram contributing to CMDM, devoting special attention to the analysis of the infrared divergence coming from the triple gluon diagram, which is addressed by four different methods leading to the same divergent behaviour.

\section{Conclusions}
\label{Sec:conclusions}

We have revisited the anomalous CMDM in the SM and demonstrated by DR the existence of an IR pole $1/\epsilon_\mathrm{IR}$ when the gluon is on-shell. The IR divergence is induced by the contribution of the non-Abelian triple gluon vertex diagram. Consequently, the perturbative CMDM must be evaluated with the off-shell gluon momentum transfer at the reference scales $q^2=\pm m_Z^2$ \cite{Choudhury:2014lna}; this choice is based on the strong coupling constant, which is perturbatively evaluated at the conventional spacelike value, $\alpha_s(-q^2=m_Z^2)=0.1179$ \cite{PDG2020,Field:1989uq,Deur:2016tte,BeiglboCk:2006lfa,Nesterenko:2016pmx}. The most important prediction of our work is the evaluation of the CMDM of the top quark in the spacelike scenario $\hat{\mu}_t(-m_Z^2)$ $=$ $-0.0224$ $-0.000925i$, whose real part coincides quite well with the recent experimental report $\hat{\mu}_t^\mathrm{Exp}$ $=$ $-0.024_{-0.009}^{+0.013}(\mathrm{stat})_{-0.011}^{+0.016}(\mathrm{syst})$ \cite{Sirunyan:2019eyu}, while our predicted imaginary quantity is an EW effect induced by the $W$ gauge boson.
Comparing the absolute values, we have
$\big|\hat{\mu}_t(-m_Z^2)\big|$ $=$ $0.0224$,
$\big|\hat{\mu}_t(m_Z^2)\big|$ $=$ $0.0298$, and
$\big|\hat{\mu}_t^\mathrm{Exp}\big|$ $=$ $0.024$. However, according to our results for the timelike scenario, our predictions for $\hat{\mu}_t(m_Z^2)$ should not be discarded since they fall within the experimental range of measurement.

From our obtained results for the top quark CMDM, we appreciate that both perturbative parameters $\alpha_s$ and $\hat{\mu}_t$ have similar behaviours: they are undetermined when $q^2\to 0$ and very well describe the strong interaction processes at the spacelike conventional scale $q^2=-m_Z^2$.

\section*{Acknowledgments}
This work has been partially supported by SNI-CONACYT and CIC-UMSNH. J.~M. thanks C\'atedras Conacyt project 1753.

\appendix

\section{Input values}
\label{appendix-Feynman-rules}

In our calculations, we have employed the SM Feynman rules given in Ref.~\cite{Quang:1998yw}, the electron unit charge $e=\sqrt{4\pi\alpha}$ and the QCD group strong coupling constant $g_s=\sqrt{4\pi\alpha_s}$.
We took input values from PDG 2020 \cite{PDG2020}: the strong coupling constant $\alpha_s(m_Z)=0.1179$, the weak-mixing angle $s_W\equiv$ $\sin{\theta_W}(m_Z)$ $=$ $\sqrt{0.23121}$,
the quark masses
$m_d = 0.00467$,
$m_s = 0.093$,
$m_b = 4.18$, and
$m_t = 172.76$ GeV,
the boson masses $m_W$=$80.379$, $m_Z$=$91.1876$, and $m_H=125.1$ GeV and the quark-mixing matrix of Cabibbo--Kobayashi--Maskawa (CKM) is
\begin{eqnarray}\label{}
V_\mathrm{CKM} &=&
\left(
\begin{array}{ccc}
|V_{ud}| & |V_{us}| & |V_{ub}| \\
|V_{cd}| & |V_{cs}| & |V_{cb}| \\
|V_{td}| & |V_{ts}| & |V_{tb}| \\
\end{array}
\right)
\nonumber\\
&=&
\left(
\begin{array}{ccc}
 0.9737 & 0.2245 & 0.00382 \\
 0.221 & 0.987 & 0.041 \\
 0.008 & 0.0388 & 1.013 \\
\end{array}
\right).
\end{eqnarray}
The fine-structure constant $\alpha(m_Z)=1/129$ is taken from \cite{Denner:2019vbn}.
In addition, the electric charges of the quarks $Q_{t}=2/3$ and the weak couplings $g_{Vt}=(3-8s_W^2)/6$, $g_{At}=1/2$.

\section{The Passarino--Veltman scalar functions}
\label{appendix-PaVe}

We follow the \texttt{FeynCalc} definitions for the scalar functions arguments. The Feynman parameterization formulas are:

\noindent i) The one-point scalar function
\begin{equation}\label{A0-formula}
A_0\big(m_0^2\big)=m_0^2\left(\Delta_\mathrm{UV}+\ln\frac{\mu ^2}{m_0^2-i\varepsilon}+1\right),
\end{equation}
\begin{eqnarray}\label{DivergenceUV}
\Delta_\mathrm{UV} &\equiv& (4\pi)^{\epsilon_\mathrm{UV}}\Gamma(\epsilon_\mathrm{UV})
\nonumber\\
&\approx& \frac{1}{\epsilon_\mathrm{UV}}-\gamma_E+\ln 4\pi \ ,
\end{eqnarray}
\begin{equation}\label{UV-pole}
\epsilon_\mathrm{UV}\equiv\epsilon=\frac{4-D}{2}\gtrsim 0;
\end{equation}
ii) The two-point scalar function
\begin{equation}\label{B0-formula}
B_0\big(q_1^2,m_0^2,m_1^2\big)
=\Delta_\mathrm{UV}+\int_0^1dx_1\ln\frac{\mu^2}{\Delta B_0} \ ,
\end{equation}
\begin{equation}\label{}
\Delta B_0\equiv q_1^2 x_1^2+(m_0^2-m_1^2-q_1^2)x_1+m_1^2-i\varepsilon \ ;
\end{equation}
iii) The three-point scalar function
\begin{eqnarray}\label{C0-formula}
&& C_0\big(q_1^2,(q_1-q_2)^2,q_2^2,m_0^2,m_1^2,m_2^3\big)=
\int_0^1dx_1 \int_0^{1-x_1}dx_2
\frac{-1}{\Delta C_0} ,
\end{eqnarray}
\begin{eqnarray}
\Delta C_0 &=&
q_2^2x_1^2+(q_1-q_2)^2x_2^2+(m_0^2-m_2^2-q_2^2)x_1
+[m_1^2-m_2^2-(q_1-q_2)^2]x_2
\nonumber\\
&& +[-q_1^2+q_2^2+(q_1-q_2)^2]x_1x_2
 +m_2^2-i\varepsilon.
\end{eqnarray}

\

Explicit solutions:
\begin{equation}\label{B0-1}
B_0(m_q^2,0,m_q^2)=\Delta_\mathrm{UV}+\ln\frac{\mu^2}{m_q^2}+2 .
\end{equation}

\begin{equation}\label{B0-2}
B_0(q^2,m_q^2,m_q^2)=
\Delta_\mathrm{UV}+\ln\frac{\mu^2}{m_q^2}+2
+\frac{R_q}{q^2} \ln\frac{R_q+2 m_q^2-q^2}{2 m_q^2},
\end{equation}
with $R_q\equiv \sqrt{q^2 \left(q^2-4 m_q^2\right)}$.

\begin{equation}\label{B0-3}
B_0(0,m_q^2,m_q^2)=
\Delta_\mathrm{UV}+\ln\frac{\mu^2}{m_q^2}.
\end{equation}

\begin{eqnarray}\label{B0-4}
B_0(m_q^2,m_q^2,m_X^2) &=&
\Delta_\mathrm{UV}+\ln\frac{\mu ^2}{m_q^2}+2
+\frac{m_X^2}{2m_q^2}\ln\frac{m_q^2}{m_X^2}
+\frac{m_X}{m_q^2}R_X\ln\frac{R_X+m_X}{2 m_q},
\end{eqnarray}
with $R_X\equiv\sqrt{m_X^2-4 m_q^2}$.

\begin{eqnarray}\label{B0-5}
B_0\left(m_{q_i}^2,m_{q_j}^2,m_W^2\right)
&=& \Delta_\mathrm{UV}  +\ln \frac{\mu ^2}{m_W^2}+2
-\frac{m_{q_i}^2+m_{q_j}^2-m_W^2}{2m_{q_i}^2} \ln \frac{m_{q_j}^2}{m_W^2}
+\frac{\sqrt{a}}{m_{q_i}^2}\ln\frac{\sqrt{a}-m_{q_i}^2+m_{q_j}^2+m_W^2}{2m_{q_j}m_W},
\nonumber\\
\end{eqnarray}
with $a\equiv m_{q_i}^4-2 m_{q_i}^2 \left(m_{q_j}^2+m_W^2\right)+\left(m_{q_j}^2-m_W^2\right)^2$.

\begin{equation}\label{B0-3g}
B_0(q^2,0,0)=\Delta_\mathrm{UV}+2+\ln\frac{~\mu^2}{-q^2}.
\end{equation}

\begin{eqnarray}\label{C0-1}
&& C_0\left(m_q^2,m_q^2,0,m_q^2,m_X^2,m_q^2\right)=
-\frac{1}{2m_q^2}
\left(
\ln\frac{m_q^2}{m_X^2}
+\frac{2m_X}{R_X}
\ln\frac{R_X+m_X}{2 m_q}
\right).
\end{eqnarray}

\begin{eqnarray}\label{C0-3}
C_0\left(m_q^2,m_q^2,q^2,m_q^2,m_X^2,m_q^2\right) &\approx&
\frac{1}{60m_q^6}
\left(
-\left[30m_q^4+\left(5m_q^2+q^2\right)q^2\right]\ln\frac{m_q^2}{m_X^2}
\right.
\nonumber\\
&&
\left.
+\frac{m_q^2q^2\left\{
4m_q^2\left[20m_q^2-\left(5m_X^2-6q^2\right)\right]-3m_X^2q^2\right\}}{\left(m_X^2-4m_q^2\right)^2}
\right.
\nonumber\\
&& \left.
-\frac{2m_X}{\left(m_X^2-4m_q^2\right)^{5/2}}
\left\{120m_q^6\left[4m_q^2-\left(2m_X^2-q^2\right)\right]
+10m_q^4\left[3m_X^4+(3q^2-5m_X^2)q^2\right]
\right.\right.
\nonumber\\
&& \left.\left. +m_X^2q^2\left[5m_q^2\left(m_X^2-2q^2\right)+m_X^2q^2\right]\right\}
\ln\frac{R_X+m_X}{2 m_q}
\right),
\nonumber\\
\end{eqnarray}
where $m_q^2>m_X^2\geq q^2$.

\begin{eqnarray}\label{C0-4}
&& C_0(m_{q_i}^2, m_{q_i}^2, 0, m_{q_j}^2, m_W^2, m_{q_j}^2) =
-\frac{1}{m_{q_i}^2}\left[
\frac{1}{2}\ln\frac{m_{q_j}^2}{m_W^2}
+\frac{m_{q_i}^2-m_{q_j}^2+m_W^2}{\sqrt{a}}
\right.
\left.
\ln\frac{\sqrt{a}-m_{q_i}^2+m_{q_j}^2+m_W^2}{2 m_{q_j} m_W}\right],
\end{eqnarray}
with $a\equiv m_{q_i}^4-2 m_{q_i}^2 \left(m_{q_j}^2+m_W^2\right)+\left(m_{q_j}^2-m_W^2\right)^2$.

\begin{eqnarray}\label{C0-3g}
C_0(m_{q_i}^2, m_{q_i}^2, q^2, 0, m_{q_i}^2, 0) &=&
\frac{1}{6q^2\sqrt{1-\frac{4 m_{q_i}^2}{q^2}}}
\left\{4\pi^2
+3 \ln ^2\left[1
+\frac{q^2}{2 m_{q_i}^2}
\left(\sqrt{1-\frac{4 m_{q_i}^2}{q^2}}-1\right)\right]
\right.
\nonumber\\
&+&
\left.
12 \mathrm{Li}_2\left[1+\frac{q^2}{2 m_{q_i}^2}
\left(\sqrt{1-\frac{4 m_{q_i}^2}{q^2}}-1\right)\right]
\right\}.
\nonumber\\
\end{eqnarray}

\section{Dimensional regularization and the IR divergence}
\label{appendix-IR-regularization}

Starting from D dimensions for the two-point scalar function $B_{02}^{3g} \equiv B_0(q^2,0,0)$, see Eq.~(\ref{B0-3g}), it is found that this function is responsible for the IR divergence when the gluon is on-shell, $q^2=0$ (this appears in Eqs. (\ref{MDM-3g}) and (\ref{IR-divergence-1})). This procedure will help us to reveal their ultraviolet $1/\epsilon_\mathrm{UV}$ and infrared $1/\epsilon_\mathrm{IR}$ poles, for which we follow Refs.~\cite{Muta:2010xua,Ilisie:2016jta}.
The integral representation that gives rise to that PaVe is
\begin{eqnarray}\label{B0-UV-IR}
B_0(q^2,0,0) &=& -i16\pi^2\mu^{2\epsilon}\int\frac{d^Dk}{(2\pi)^D}\frac{1}{(k-p')^2(k-p)^2}
\nonumber\\
&=& -i16\pi^2\mu^{2\epsilon}\int\frac{d^Dk}{(2\pi)^D}\frac{1}{k^2(k+q)^2}.
\end{eqnarray}

To regularize $B_0(q^2,0,0)$ when $q^2\to0$, we use\footnote{\label{epsilon-Ilisie} In Ref.~\cite{Ilisie:2016jta}, $D=4+2\epsilon$ is used, where $\epsilon\lesssim0$ stands for the UV divergence and $\epsilon\gtrsim0$ for the IR divergence.}
the space-time dimension $D=4-2\epsilon$ as in Refs. \cite{Muta:2010xua,Denner:2019vbn}, being $\epsilon_\mathrm{UV}\equiv\epsilon\gtrsim0$ for the UV divergence and $\epsilon_\mathrm{IR}\equiv\epsilon\lesssim0$ for the IR divergence.
The FP for the integrand in Eq.~(\ref{B0-UV-IR}) is
\begin{eqnarray}\label{}
\frac{1}{k^2(k+q)^2} &=& \int_0^1dx_1\frac{\Gamma(2)}{\left[x_1k^2+(1-x_1)(k+q)^2\right]^2}
\nonumber\\
&=& \int_0^1dx_1\frac{\Gamma(2)}{(\ell^2-\Delta B_0)^2}~,
\end{eqnarray}
with $k\equiv \ell+q(x_1-1)$, $dk=d\ell$ and
$\Delta B_0\equiv -q^2x_1(1-x_1)$. Using the $D$-dimensional Minkowski integral given in Eq.~(A.44) from \cite{Peskin:1995ev}
\begin{equation}\label{}
\int\frac{d^D\ell}{(2\pi)^D}\frac{1}{(\ell^2-\Delta)^n} =
\frac{(-1)^ni}{(4\pi)^{D/2}}
\frac{\Gamma(n-\frac{D}{2})}{\Gamma(n)}\frac{1}{~\Delta^{n-D/2}}~,
\end{equation}
and the Euler beta function
\begin{eqnarray}\label{}
\mathcal{B}(x,y) &=& \int_0^1dz~z^{x-1}(1-z)^{y-1}
\nonumber\\
&=& \frac{\Gamma(x)\Gamma(y)}{\Gamma(x+y)}~,
\end{eqnarray}
we obtain
\begin{eqnarray}\label{B0-regularized-stage1}
B_0(q^2,0,0) &=& -i16\pi^2\mu^{2\epsilon}\int\frac{d^D\ell}{(2\pi)^D}
\int_0^1dx_1\frac{\Gamma(2)}{(\ell^2-\Delta B_0)^2}
\nonumber\\
&=& \Gamma(\epsilon)\int_0^1dx_1\left(\frac{4\pi\mu^2}{\Delta B_0}\right)^\epsilon
\nonumber\\
&=& \Gamma(\epsilon)\left(\frac{4\pi\mu^2}{-q^2}\right)^\epsilon
\int_0^1dx_1\frac{1}{x_1^\epsilon(1-x_1)^\epsilon}
\nonumber\\
&=& \Gamma(\epsilon)
\left(\frac{4\pi\mu^2}{-q^2}\right)^\epsilon
\mathcal{B}(1-\epsilon,1-\epsilon)
\nonumber\\
&=& \Gamma(\epsilon)\left(\frac{4\pi\mu^2}{-q^2}\right)^\epsilon,
\end{eqnarray}
where $\mathcal{B}(1-\epsilon,1-\epsilon)=1$ for $\epsilon\to 0$. In addition, the term $1/(-q^2)^\epsilon$ with $\epsilon\to 0$ and $q^2\to 0$ is an indetermination of the type 1/$0^0$, where the procedure of taking the limit must be carefully performed; this problem can be faced following the method from Ref.~\cite{Ilisie:2016jta}.
First, according to the spirit of Eq.~(8.22) from Ref. \cite{Ilisie:2016jta}, we have that
\begin{equation}\label{regularization-integral-1}
\frac{1}{(-y)^\epsilon}=\epsilon\int_{-y}^\infty \frac{dx}{x} \frac{1}{x^\epsilon}, \quad
\mathrm{Re}~\epsilon\gtrsim0~,
\end{equation}
which then splits into two regions $\int_{-y}^\infty=\int_a^\infty+\int_{-y}^a$~, and hence, $\left(\frac{4\pi\mu^2}{-q^2}\right)^\epsilon$ from Eq.~(\ref{B0-regularized-stage1}) results in
\begin{eqnarray}\label{regularization-2}
\left(\frac{4\pi\mu^2}{-q^2}\right)^\epsilon &=& ~~\epsilon\int_{-q^2}^\infty
\frac{dr^2}{r^2}\left(\frac{4\pi\mu^2}{r^2}\right)^{\epsilon}
\nonumber\\
&=&~~\epsilon\int_{4\pi\mu^2}^\infty
\frac{dr^2}{r^2}\left(\frac{4\pi\mu^2}{r^2}\right)^{\epsilon}
+\epsilon\int_{-q^2}^{4\pi\mu^2}
\frac{dr^2}{r^2}\left(\frac{4\pi\mu^2}{r^2}\right)^{\epsilon}.
\end{eqnarray}
In the same context of Eqs.~(8.24) from Ref. \cite{Ilisie:2016jta}, it can be written that:

\noindent UV region
\begin{equation}\label{regularization-integral-2}
\int_{a}^\infty \frac{dx}{x}\left(\frac{a}{x}\right)^\epsilon
=\frac{1}{\epsilon}\equiv~\frac{1}{\epsilon_\mathrm{UV}}, \quad
\mathrm{Re}~\epsilon\gtrsim0, \
a > 0  ;
\end{equation}
IR region
\begin{equation}\label{regularization-integral-3}
\int_0^a \frac{dx}{x}\left(\frac{a}{x}\right)^\epsilon=-\frac{1}{\epsilon}\equiv-\frac{1}{\epsilon_\mathrm{IR}}, \quad
\mathrm{Re}~\epsilon\lesssim0 , \
a > 0 .
\end{equation}
Then, by replacing Eq.~(\ref{regularization-2}) in Eq.~(\ref{B0-regularized-stage1}) and performing $q^2\to 0$ by means of Eqs.~(\ref{regularization-integral-2}) and (\ref{regularization-integral-3}), we obtain
\begin{eqnarray}
B_0(q^2,0,0) &=& \Gamma(\epsilon)
~\epsilon\int_{4\pi\mu^2}^\infty \frac{dr^2}{r^2}\left(\frac{4\pi\mu^2}{r^2}\right)^{\epsilon}
+
\Gamma(\epsilon)
~\epsilon\int_{-q^2}^{4\pi\mu^2}\frac{dr^2}{r^2}\left(\frac{4\pi\mu^2}{r^2}\right)^{\epsilon},
\end{eqnarray}
and taking the limit
\begin{eqnarray}
\lim_{q^2\to 0} B_0(q^2,0,0)&=&
\Gamma(\epsilon_\mathrm{UV})
-
\Gamma(\epsilon_\mathrm{IR})
\nonumber\\
&\approx&
\frac{1}{\epsilon_\mathrm{UV}}-\frac{1}{\epsilon_\mathrm{IR}}
\nonumber\\
&=& B_0(0,0,0)~,
\end{eqnarray}
where $\Gamma(\epsilon_\mathrm{UV})\approx 1/\epsilon_\mathrm{UV}-\gamma_E$ and
$\Gamma(\epsilon_\mathrm{IR})\approx 1/\epsilon_\mathrm{IR}-\gamma_E$, it can be expressed as
\begin{eqnarray}\label{B0-3g-final}
B_0(0,0,0) &=& \frac{1}{\epsilon_\mathrm{UV}}-\frac{1}{\epsilon_\mathrm{IR}}
\nonumber\\
&=&
\Delta_\mathrm{UV}-\Delta_\mathrm{IR}~.
\end{eqnarray}
Now, the UV and IR poles are explicit, being suitable to express the poles through $\Delta_\mathrm{UV}$ from Eq.~(\ref{DivergenceUV}) and an analogous definition for the IR divergence:
\begin{eqnarray}\label{DivergenceIR}
\Delta_\mathrm{IR}&\equiv&(4\pi)^{\epsilon_\mathrm{IR}}\Gamma(\epsilon_\mathrm{IR})
\nonumber\\
&\approx& \frac{1}{\epsilon_\mathrm{IR}}-\gamma_E+\ln 4\pi,
\end{eqnarray}
\begin{equation}
\epsilon_\mathrm{IR}\equiv\epsilon=\frac{4-D}{2}\lesssim0~.
\end{equation}

Finally, with $B_{01}^{3g} \equiv B_0(m_{q_i}^2,0,m_{q_i}^2)$ from Eq.~(\ref{B0-1}) and the new
$B_{02}^{3g} \equiv B_0(0,0,0)$ from Eq.~(\ref{B0-3g-final}) for $q^2=0$, the on-shell case of Eq.~(\ref{IR-divergence-1}) takes the form
\begin{eqnarray}
B_{01}^{3g}-B_{02}^{3g}
&=&B_0(m_{q_i}^2,0,m_{q_i}^2)-B_0(0,0,0)
\nonumber\\
&=&\Delta_\mathrm{IR}+\ln\frac{\mu^2}{m_{q_i}^2}+2,
\end{eqnarray}
which exhibits the IR nature of the divergence contained in the CMDM of the triple gluon vertex diagram.

\section{Feynman parameterization of the triple gluon vertex diagram with the $+i\varepsilon$ prescription}
\label{appendix-3g-parameterizations}

The triple gluon vertex contribution to the CMDM (see Fig.~\ref{FIGURE-chromo}(f)), with the on-shell gluon, was calculated in the SM in Refs.~\cite{Choudhury:2014lna,Martinez:2007qf} by means of the FP method. Nonetheless, the authors of Ref. \cite{Martinez:2007qf} did not carefully consider the $+i\varepsilon$ Feynman prescription for the propagator~\cite{Peskin:1995ev,Halzen:1984mc,Davydychev:1992xr,Zee:2003mt}, whilst in Ref.~\cite{Choudhury:2014lna}, even if the IR divergence was indicated and the artifice of massive gluons was also implemented, the authors did not provide analytical solutions for the parameterized integrals.

In what follows, we demonstrate analytically through the FP method, with the strict use of the $+i\varepsilon$ prescription of the propagators, that $\hat{\mu}_{q_i}(3g)$ with the on-shell gluon generates a logarithmic IR divergence. By keeping $+i\varepsilon$ throughout the calculation, once the parameterized integrals have been completely solved, we can now apply the limit $\varepsilon\to 0$; this is implemented in order to find out whether the solution is finite or not. In consequence, only after knowing that the solution is truly finite is it correct to set $\varepsilon=0$ from the beginning. In the case of $\hat{\mu}_{q_i}(3g)$, a logarithmic divergence arises when $\varepsilon\to 0$. We will approach this problem in two different ways: in Sect. \ref{appendix-3g-FP-xi1}, the gluon propagator in the Feynman-'t Hooft gauge ($\xi=1$) is considered, and in Sect. \ref{appendix-3g-FP-Rxi}, the gluon propagator in the general renormalizable $R_\xi$ gauge is implemented; this is the case in order to know whether or not there is dependence on the $\xi$ gauge parameter. Furthermore, the artifice of the massive gluon propagator is carried out in Sect. \ref{appendix-3g-artifice}.

\subsection{The gluon propagator in the Feynman-'t Hooft gauge}
\label{appendix-3g-FP-xi1}

The gluon propagator in the general $R_\xi$ gauge is
\begin{equation}\label{gluon-propagator-Rxi}
\frac{i}{p^2+i\varepsilon}\left[-g_{\mu\nu}+(1-\xi)\frac{p_\mu p_\nu}{p^2+i\varepsilon}\right]\delta_{ab}.
\end{equation}
This propagator in the Feynman-'t Hooft gauge $\xi=1$ was used in the integral from Eq.~(\ref{3g-integral}) in order to obtain $\hat{\mu}_{q_i}(3g)$ in Eq.~(\ref{MDM-3g}) with the off-shell gluon.

In the following, we compute $\hat{\mu}_{q_i}(3g)$ when $\xi=1$, first for the off-shell gluon and after with the on-shell gluon.

i) The off-shell gluon case ($q^2\neq 0$). From (\ref{3g-integral}), with its denominator parameterized via

\begin{eqnarray}\label{parameterization-simple}
\frac{1}{D_1D_2D_3}&=&
\int_0^1dx_1\int_0^{1-x_1}dx_2
\frac{\Gamma(3)}{\left[x_1D_1+x_2D_2+(1-x_1-x_2)D_3\right]^3},
\end{eqnarray}
where $D_1\equiv k^2-m_{q_i}^2+i\varepsilon$,
$D_2\equiv (k-p')^2+i\varepsilon$,
$D_3\equiv (k-p)^2+i\varepsilon$
and the shift $k=\ell-p~x_1+(p'-p)x_2+p$, the resulting CMDM is
\begin{eqnarray}\label{3g-parameterization}
\hat{\mu}_{q_i}(3g)_{\xi=1} &=&
\frac{3\alpha_s}{4\pi}
\int_0^1dx_1\int_0^{1-x_1}dx_2
\frac{m_{q_i}^2x_1(x_1-1)}{m_{q_i}^2x_1^2+q^2(x_1+x_2-1)x_2-i\varepsilon}.
\end{eqnarray}
This reproduces the same numerical results as those obtained with Eq.~(\ref{MDM-3g}), where the Passarino--Veltman tensor decomposition method was employed. We have implemented high numerical precision to successfully evaluate (\ref{3g-parameterization}) via \texttt{Mathematica}.

ii) The on-shell gluon case ($q^2=0$). Note that if $i\varepsilon$ is ignored in Eq.~(\ref{3g-parameterization}), then
\begin{eqnarray}\label{}
\hat{\mu}_{q_i}(3g)_{\xi=1} &=& -\frac{3\alpha_s}{4\pi}
\int_0^1dx_1\frac{(1-x_1)^2}{x_1}
\nonumber\\
&=& \left. -\frac{3\alpha_s}{8\pi} \left[(x_1-4)x_1+\ln x_1^2\right] \right|_0^1,
\end{eqnarray}
which diverges when $x_1\to 0$; this behaviour was pointed out in Eq.~(11) from Ref. \cite{Choudhury:2014lna}. However, we prefer to utilize Eq.~(\ref{3g-parameterization}), keeping the $+i\varepsilon$ prescription, since this allows us to entirely solve the integral and, at the end, to analyse it when $\varepsilon\to 0$. Thereby,
\begin{eqnarray}\label{3g-parameterization-xi1-qq0}
\hat{\mu}_{q_i}(3g)_{\xi=1} &=& \frac{3\alpha_s}{16\pi} \left\{
6-\left(1+\frac{i\varepsilon}{m_{q_i}^2}\right)
\ln\left(1-\frac{m_{q_i}^2}{i\varepsilon}\right)^2
\right.
\left.
-\sqrt{\frac{i\varepsilon}{m_{q_i}^2}}
\ln\left(\frac{\sqrt{i\varepsilon/m_{q_i}^2}+1}{\sqrt{i\varepsilon/m_{q_i}^2}-1}\right)^4
\right\}
\nonumber\\
&\approx& \frac{3\alpha_s}{8\pi}\left(3-\ln\frac{~m_{q_i}^2}{-i\varepsilon}\right),
\quad 0<\varepsilon\ll 1
\end{eqnarray}
which diverges when $\varepsilon\to0$.

\subsection{The gluon propagator in the general renormalizable $R_\xi$ gauge}
\label{appendix-3g-FP-Rxi}

By considering the gluon propagator in the $R_\xi$ gauge (see Eq.~(\ref{gluon-propagator-Rxi})), the integral in Eq.~(\ref{3g-integral}) takes a more complicated form:
\begin{eqnarray}\label{3g-integral-Rxi}
\mathcal{M}_{q_i}^\mu(3g) &=& \int\frac{d^Dk}{(2\pi)^D}\bar{u}(p')\left(-ig_s\gamma^{\alpha_1}T_{AC_1}^{a_1}\right)
\left(i\frac{\slashed{k}+m_{q_i}}{k^2-m_{q_i}^2+i\varepsilon}\delta_{C_1C_2}\right)\left(-ig_s\gamma^{\alpha_4}T_{C_2B}^{a_4}\right)u(p)
\nonumber\\
&&\times
\left\{\frac{i}{(k-p')^2+i\varepsilon}
\left[-g_{\alpha_1\alpha_2}+
(1-\xi)
\frac{(k-p')_{\alpha_1}(k-p')_{\alpha_2}}{(k-p')^2+i\varepsilon}\right]
\delta_{a_1a_2}\right\}
\nonumber\\
&&\times
\left[-g_sf_{aa_3a_2}T^{\mu\alpha_3\alpha_2}_{ggg}(p'-p,-k+p,k-p')\right]
\nonumber\\
&&\times
\left\{\frac{i}{(k-p)^2+i\varepsilon}
\left[-g_{\alpha_3\alpha_4}+(1-\xi)
\frac{(k-p)_{\alpha_3}(k-p)_{\alpha_4}}{(k-p)^2+i\varepsilon}\right]
\delta_{a_3a_4}\right\}
\nonumber\\
&=&-\frac{3g_s^3}{2}T_{AB}^a
\int\frac{d^Dk}{(2\pi)^D}
\frac{\bar{u}(p')\gamma^{\alpha_1}(\slashed{k}+m_{q_i})\gamma^{\alpha_4}u(p)}
{\left(k^2-m_{q_i}^2+i\varepsilon\right)\left[(k-p')^2+i\varepsilon\right]^2\left[(k-p)^2+i\varepsilon\right]^2}
\nonumber\\
&&\times
\left\{-g_{\alpha_1\alpha_2}\left[(k-p')^2+i\varepsilon\right]+
(1-\xi)
(k-p')_{\alpha_1}(k-p')_{\alpha_2}\right\}
\nonumber\\
&&\times ~
T^{\mu\alpha_3\alpha_2}_{ggg}(p'-p,-k+p,k-p')
\nonumber\\
&&\times
\left\{-g_{\alpha_3\alpha_4}\left[(k-p)^2+i\varepsilon\right]+
(1-\xi)
(k-p)_{\alpha_3}(k-p)_{\alpha_4}\right\},
\end{eqnarray}
whose denominator is parameterized as
\begin{eqnarray}\label{}
\frac{1}{D_1D_2^2D_3^2} &=& \frac{\Gamma(5)}{\Gamma(1)\Gamma(2)\Gamma(2)}
\int_0^1dx_1\int_0^{1-x_1}dx_2
\frac{x_2(1-x_1-x_2)}{\left[x_1D_1+x_2D_2+(1-x_1-x_2)D_3\right]^5}~,
\nonumber\\
\end{eqnarray}
where $D_{1,2,3}$ and the shift of $k$ are the same as in Eq.~(\ref{parameterization-simple}). Thus, the resulting CMDM is
\begin{eqnarray}\label{3g-parameterization-Rxi-qq0}
\hat{\mu}_{q_i}(3g)_{R_\xi} &&= \frac{\alpha_s}{32\pi}
\int_0^1dx_1\int_0^{1-x_1}dx_2
\left\{
\frac{4 m_{q_i}^2 (5 x_1-3) x_1}{m_{q_i}^2x_1^2+q^2(x_1+x_2-1)x_2-i\varepsilon}
+\frac{2(1-\xi)m_{q_i}^2(5 x_1-2) x_1}{m_{q_i}^2x_1^2+q^2(x_1+x_2-1)x_2-i\varepsilon}
\right.
\nonumber\\
&& \left. +\frac{
2 m_{q_i}^2 \left[m_{q_i}^2 x_1^3 (4-5 x_1)+ i\varepsilon x_1 (3-4 x_1) \right]}
{\left[m_{q_i}^2x_1^2+q^2(x_1+x_2-1)x_2-i\varepsilon\right]^2}
+\frac{(1-\xi) m_{q_i}^2 \left[m_{q_i}^2 x_1^3 (6-5 x_1)+i\varepsilon x_1 (1-2 x_1) \right]
}{\left[m_{q_i}^2x_1^2+q^2(x_1+x_2-1)x_2-i\varepsilon\right]^2}
\right.
\nonumber\\
&& \left.
+\frac{2 (x_1-1) x_1 \left(m_{q_i}^3 x_1^2+i\varepsilon m_{q_i} \right)^2}
{\left[m_{q_i}^2x_1^2+q^2(x_1+x_2-1)x_2-i\varepsilon\right]^3}
+\frac{(1-\xi)
 m_{q_i}^4(x_1-2) x_1^3 \left(m_{q_i}^2 x_1^2+i \varepsilon \right)}{\left[m_{q_i}^2x_1^2+q^2(x_1+x_2-1)x_2-i\varepsilon\right]^3} \right\}.
\end{eqnarray}
Solving $\hat{\mu}_{q_i}(3g)_{R_\xi}$ for $q^2=0$ yields
\begin{equation}\label{}
\hat{\mu}_{q_i}(3g)_{R_\xi}=\hat{\mu}_{q_i}(3g)_{\xi=1}+\hat{\mu}_{q_i}(3g)_{\xi}~,
\end{equation}
which is divergent since the first term, $\hat{\mu}_{q_i}(3g)_{\xi=1}$, is the same as that obtained with the Feynman-'t Hooft gauge in Eq.~(\ref{3g-parameterization-xi1-qq0}); it should be recalled that this diverges when $\varepsilon\to0$. On the other hand, the second term, $\hat{\mu}_{q_i}(3g)_{\xi}$, is proportional to the $\xi$ gauge parameter as
\begin{eqnarray}\label{3g-parameterization-xi-qq0}
\hat{\mu}_{q_i}(3g)_{\xi} &=& \frac{3\alpha_s}{64\pi}(1-\xi) \left\{
2\frac{i\varepsilon}{m_{q_i}^2}
\left[
3-\ln\left(1-\frac{m_{q_i}^2}{i\varepsilon}\right)^2
\right]
\right.
\left.
-\sqrt{\frac{i\varepsilon}{m_{q_i}^2}}\left(1+3\frac{i\varepsilon}{m_{q_i}^2}\right)
\ln\frac{\sqrt{i\varepsilon/m_{q_i}^2}+1}{\sqrt{i\varepsilon/m_{q_i}^2}-1}
\right\}
\nonumber\\
&\approx& \frac{i3\alpha_s}{64}(1-\xi)\sqrt{\frac{i\varepsilon}{m_{q_i}^2}}, \quad 0<\varepsilon\ll 1
\end{eqnarray}
but it vanishes when $\varepsilon\to0$; thus, it is proven that $\hat{\mu}_{q_i}(3g)_{R_\xi}$ is independent of the $\xi$ gauge parameter.

\subsection{The massive gluon propagator artifice}
\label{appendix-3g-artifice}

Another way to address the IR divergence is providing small fictitious masses in the gluon propagator. This method was used in Ref.~\cite{Choudhury:2014lna} to show numerically that the triple gluon vertex contribution to the CMDM for $q^2=0$ diverges when $m_g\to 0$. Applying such artifice to the loop integral in Eq.~(\ref{3g-integral}), it takes the form
\begin{eqnarray}\label{3g-integral-mg}
\mathcal{M}_{q_i}^\mu(3g) &=& -\frac{3g_s^3}{2}T_{AB}^a
\int\frac{d^Dk}{(2\pi)^D}
\frac{\bar{u}(p')\gamma^{\alpha_1}(\slashed{k}+m_{q_i})\gamma^{\alpha_4}u(p)}
{\left(k^2-m_{q_i}^2+i\varepsilon\right)\left[(k-p')^2-m_g^2+i\varepsilon\right]\left[(k-p)^2-m_g^2+i\varepsilon\right]}
\nonumber\\
&&\times
\left(-g_{\alpha_1\alpha_2}\right)
T^{\mu\alpha_3\alpha_2}_{ggg}(p'-p,-k+p,k-p')
\left(-g_{\alpha_3\alpha_4}\right),
\end{eqnarray}
here $+i\varepsilon$ can be omitted, since taking the limit $\varepsilon\to 0$ will ultimately lead to a finite solution as long as $m_g\neq 0$; we will keep it by formality. The starting point is to parameterize the denominator as in Eq.~(\ref{parameterization-simple}), with the same shift for $k$ but with $D_1\equiv k^2-m_{q_i}^2+i\varepsilon$,
$D_2\equiv (k-p')^2-m_g^2+i\varepsilon$ and $D_3\equiv (k-p)^2-m_g^2+i\varepsilon$.
The derived CMDM with the on-shell gluon is
\begin{eqnarray}\label{3g-parameterization-artifice}
\hat{\mu}_{q_i}(3g)_{m_g} &=& \frac{3\alpha_s}{4\pi}
\int_0^1dx_1\int_0^{1-x_1}dx_2
\frac{m_{q_i}^2x_1(x_1-1)}{m_{q_i}^2 x_1^2-m_g^2(x_1-1)-i\varepsilon}
\nonumber\\
&=&\frac{3\alpha_s}{8\pi}\left\{
3-2\frac{m_g^2}{m_{q_i}^2}+\left(1-3\frac{m_g^2}{m_{q_i}^2}+\frac{m_g^4}{m_{q_i}^4}
+\frac{i\varepsilon}{m_{q_i}^2}\right)
\ln\frac{m_g^2-i\varepsilon}{m_{q_i}^2-i\varepsilon}
\right\}
\nonumber\\
&\approx& \frac{3\alpha_s}{8\pi}\left(3-\ln\frac{m_{q_i}^2}{m_g^2}\right), \quad 0<m_g\ll 1
\end{eqnarray}
which also diverges when $m_g\to 0$. Notice that this final expression is independent of the $\varepsilon$ parameter; this is why $\varepsilon$ can be removed at the outset.

\end{document}